\documentclass[prd,preprint,preprintnumbers,nofootinbib,eqsecnum,superscriptaddress]{revtex4}

 \usepackage[dvips,final]{graphicx}
  \usepackage{amssymb}
   \usepackage{amsmath}
    \usepackage{amsfonts}
     \usepackage{epsfig}
      \usepackage{bm}% bold math

\usepackage{mathpazo}

\usepackage[section]{placeins}

\usepackage{ctable}
\usepackage{booktabs}
\usepackage{array}
\usepackage{tabularx}
\usepackage{xcolor}
\usepackage{pstricks}
\definecolor{fred}{rgb}{0.90053, 0.00369, 0.00159}  % ta3skyblue

\newcommand{\bp}{\mbox{\boldmath $p$}}
\newcommand{\bq}{\mbox{\boldmath $q$}}

\begin{document}

\author{Antoni Szczurek\footnote{also at University of Rzesz\'ow, 
PL-35-959 Rzesz\'ow, Poland}}
\email{antoni.szczurek@ifj.edu.pl} 
\affiliation{Institute of Nuclear
Physics, Polish Academy of Sciences, ul. Radzikowskiego 152, PL-31-342 Krak{\'o}w, Poland}

\author{Barbara Linek}
\email{basialinek@gmail.com}
\affiliation{College of Natural Sciences, Institute of Physics,
University of Rzesz\'ow, ul. Pigonia 1, PL-35-959 Rzesz\'ow, Poland}

\author{Marta {\L}uszczak}
\email{luszak@ur.edu.pl}
\affiliation{College of Natural Sciences, Institute of Physics,
University of Rzesz\'ow, ul. Pigonia 1, PL-35-959 Rzesz\'ow, Poland}

\title{Semiexclusive dilepton production in proton-proton
collisions with one forward proton measurement at the LHC}

\begin{abstract}
We discuss photon-photon fusion mechanisms of dilepton production 
in proton-proton collisions with rapidity gap in the main detector 
and one forward proton in the forward proton detectors. 
This is relevant for the LHC measurements by ATLAS+AFP and CMS+PPS. 
Transverse momenta of the intermediate photons are taken into account
and photon fluxes are expressed in terms of
proton electromagnetic form factors and structure functions.
Differential distributions in
$\xi_{1/2}$, $M_{ll}$, $Y_{ll}$, $p_{t,ll}$, $M_R$
%pamietac wyjasnic M_R w tekscie
are shown 
and the competition of different mechanisms is discussed. 
Both double-elastic and single-dissociative processes are included
in the analysis. Different parametrizations of the structure functions
are used.
We discuss also mechanism with one forward $\Delta^+$ isobar, 
or other proton resonances in the final state.
The role of several cuts is studied.
We also use SuperChic generator and compare corresponding results 
to the results of our codes.
The soft rapidity gap survival factor is calculated for each contribution
separately. The gap survival factor for the single-dissociative mechanism
due to minijet emission into the main detector is calculated in addition.
It depends on the type of contribution (fully elastic, single dissociation,
double dissociation). The soft rapidity gap survival factor for the case
of single proton measurement is significantly smaller than that for 
the inclusive case (no proton measurement).
We find only weak dependence on the invariant mass of the dilepton
system as well as the lepton pair transverse momentum
and sizeable dependence on the pair rapidity.
The latter effect is rather difficult to identify experimentally.
%and the mass of the proton remnant for the single dissociation. \\
%...................................................................
\end{abstract}

\maketitle

%---------------------------
\section{Introduction}
%---------------------------

The production of dilepton pairs via photon-photon fusion was
studied both experimentally \cite{ATLAS,CMS} and theoretically.

In \cite{LSS2016} we proposed how to include transverse momenta
of fusing virtual photons. The formalism was used to calculate
distributions of several observables related to leptons.

Till recently forward going protons were not measured.
However, recently both CMS+TOTEM \cite{CMS} and ATLAS \cite{ATLAS}
measured the cases with one proton in PPS or AFP.
This automatically selects fully exclusive process or processes
with single proton dissociation.
The CMS collaboration measured only a few events as $p_t >$ 50 GeV
cut was imposed there, whereas the ATLAS colaboration had $p_t >$ 15 GeV
cut. So the ATLAS collaboration could obtain even some distributions.

Only recently the CMS collaboration \cite{CMS} and very recently 
the ATLAS collaboration \cite{ATLAS} presented results with at least 
one proton measured in forward direction. The experimental aparatus 
allows to measure only very forward protons.
In theoretical calculations one has to impose experimental limits on 
so-called $\xi$-variables (longitudinal momentum fraction loss) \cite{CMS,ATLAS}.
The limited acceptance of the forward detectors causes that 
the cross section is considerably  reduced compared to the case when 
only leptons are measured, as will be discussed here.

Here, we use the formalism developed in \cite{LSS2016,LSS2018}, which 
allows to calculate the cross section differential also in $M_X$ or $M_Y$, 
masses of the excited proton remnants.
%We shall use the formalism also here.
In \cite{FLSS2019,LFSS2019} it was discussed
%we discussed 
how to calculate gap survival
factor which is related to emission of (mini)jets produced in a DIS
process associted with $W^+ W^-$ and $t \bar t$ production, respectively. 
We shall repeat such a calculation also here for $\mu^+ \mu^-$ production.
The {\bf absorption} for double-elastic contribution was studied 
e.g. in \cite{LS2015,LS2018} using the momentum space formalism.
The impact parameter approach can be found e.g. in \cite{DS2015}.

The same processes and a similar formalism were implemented in 
the recent version of the SuperChic 4 generator \cite{HTKR2020}.
This code generates events with four momenta of outgoing particles
(leptons, protons, jets).
The authors of SuperChic implemented also soft absorption effects
in the form of kinematics-dependent gap survival factor.
In the present study we shall use also the SuperChic 4 code for
comparison and in order to estimate the soft rapidity gap 
survival probability.

%--------------------------------------------
\section{Sketch of the formalism}
%--------------------------------------------

\begin{figure}
\includegraphics[width=5cm]{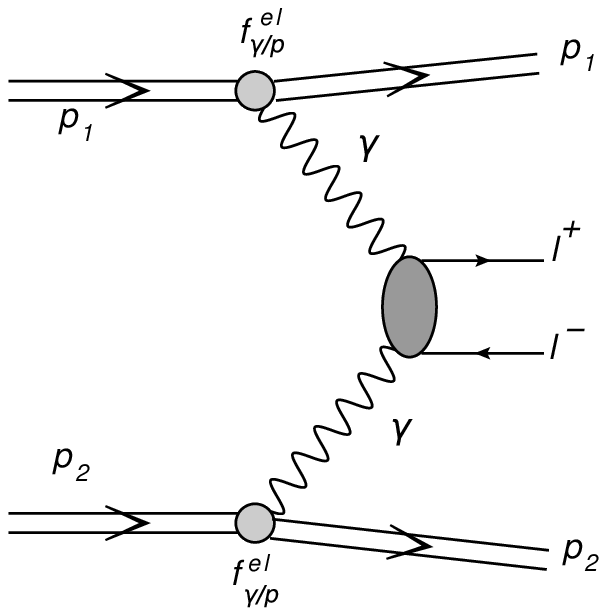}
\includegraphics[width=5cm]{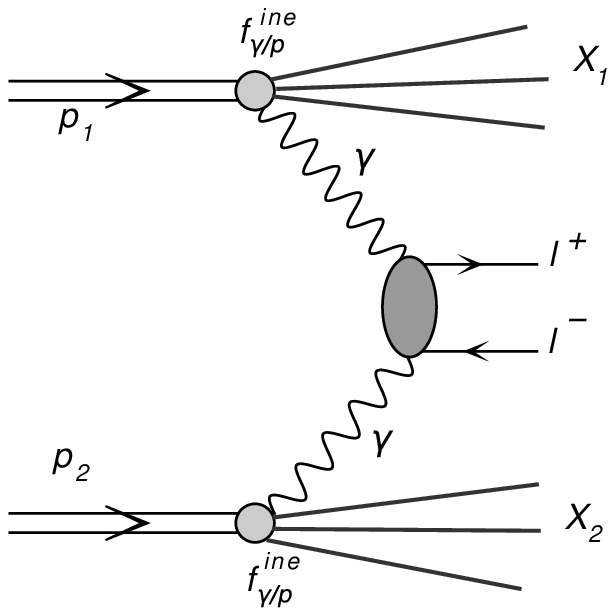}\\
\includegraphics[width=5cm]{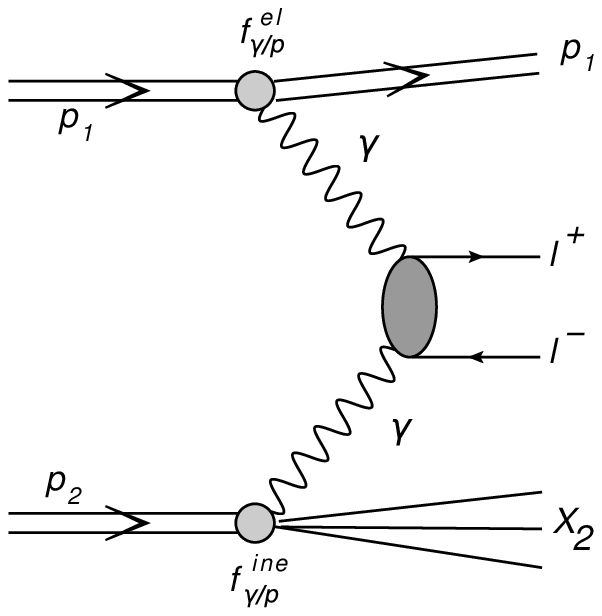}
\includegraphics[width=5cm]{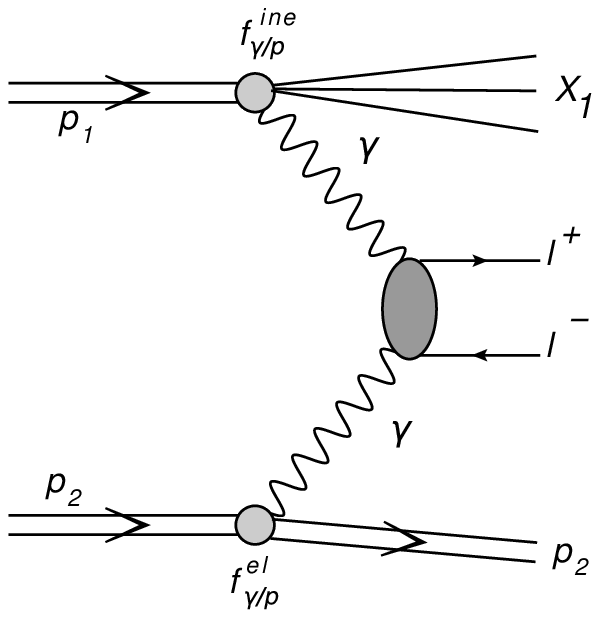}
\caption{Four different categories of $\gamma \gamma$ fusion mechanisms 
of dilepton production in proton-proton collisions.}
\label{fig:diagrams}
\end{figure}

In general, there are four categories of the $\gamma \gamma$ processes
as shown in Fig.\ref{fig:diagrams}. We shall call them
elastic-elastic, inelastic-inelastic, elastic-inelastic and
inelastic-elastic.
The first one will be also called double-elastic and the second one 
double-inelastic for brevity.

%-------------------------------------------------------------
%-------------------------------------------------------------

In the $k_T$-factorization approach  \cite{SFPSS2015,LSS2016}, the cross section
for production of $l^+l^-$ can be written in the form
\begin{eqnarray}
{d \sigma^{(i,j)} \over dy_1 dy_2 d^2\bp_1 d^2\bp_2} &&=  \int  {d^2 \bq_1 \over \pi \bq_1^2} {d^2 \bq_2 \over \pi \bq_2^2}  
{\cal{F}}^{(i)}_{\gamma^*/A}(x_1,\bq_1) \, {\cal{F}}^{(j)}_{\gamma^*/B}(x_2,\bq_2) 
{d \sigma^*(p_1,p_2;\bq_1,\bq_2) \over dy_1 dy_2 d^2\bp_1 d^2\bp_2} \, , \nonumber \\ 
\label{eq:kt-fact}
\end{eqnarray}
where the indices $i,j \in \{\rm{el}, \rm{in} \}$ denote elastic or 
inelastic final states.
Here the photon flux for {\bf inelastic} case is integrated over the mass
of the remnant.

The longitudinal momentum fractions of photons are obtained from 
the rapidities and transverse momenta of final state $l^+l^-$ as:
\begin{eqnarray}
x_1 &=& \sqrt{ {\bp_1^2 + m_l^{2} \over s}} e^{+y_1} +  
        \sqrt{ {\bp_2^2 + m_l^{2} \over s}} e^{+y_2} 
\; , \nonumber \\
x_2 &=& \sqrt{ {\bp_1^2 + m_l^{2} \over s}} e^{-y_1} 
     +  \sqrt{ {\bp_2^2 + m_l^{2} \over s}} e^{-y_2} \, .
\end{eqnarray}
The integrated fluxes for elastic and inelastic processes can be found 
in \cite{LSS2016,LSS2018}.

Then the four-momenta of intermediate photons can be written as:
\begin{eqnarray}
q_1 &\approx& \left( x_1 \frac{\sqrt{s}}{2}, \vec{q}_{1t}, x_1 \frac{\sqrt{s}}{2} \right)
\; , \nonumber \\
q_2 &\approx& \left( x_2 \frac{\sqrt{s}}{2}, \vec{q}_{2t}, -x_2 \frac{\sqrt{s}}{2} \right)
\, .
\end{eqnarray}

If one is interested in modelling what happens with the proton remnant
then the formalism must be somewhat extended.
Then the unintegrated inelastic photon distribution (flux) can be
written as:
\begin{equation}
{\cal F}_{ine}(x,q_t^2) = \int d M^2  
\frac{d {\cal F}_{ine}}{d M^2}(x,q_t^2,M^2) \; ,
\label{doubly_unintegrated_photon}
\end{equation}
where $\frac{d {\cal F}_{ine}}{d M^2}(x,q_t^2,M^2)$ is a more
differential photon distribution in the proton.
In the following we shall call it {\it doubly-unintegrated} photon 
distribution (flux).
The latter distribution was used to calculate
differential distributions for production of $W^+ W^-$ \cite{FLSS2019}
or $t \bar t$ \cite{LFSS2019} pairs with rapidity gap at midrapidities.

In principle, proton can be emitted also from the remnant system.
This requires modeling of remnant fragmentation which is not fully
under control. Such protons carry typically much reduced longitudinal
momentum fraction $x_i$ such that $\xi_i = 1 - x_i >$ 0.1,
i.e. cannot be measured in the Roman pots of the ATLAS or CMS
experiments.

\begin{figure}
\includegraphics[width=6cm]{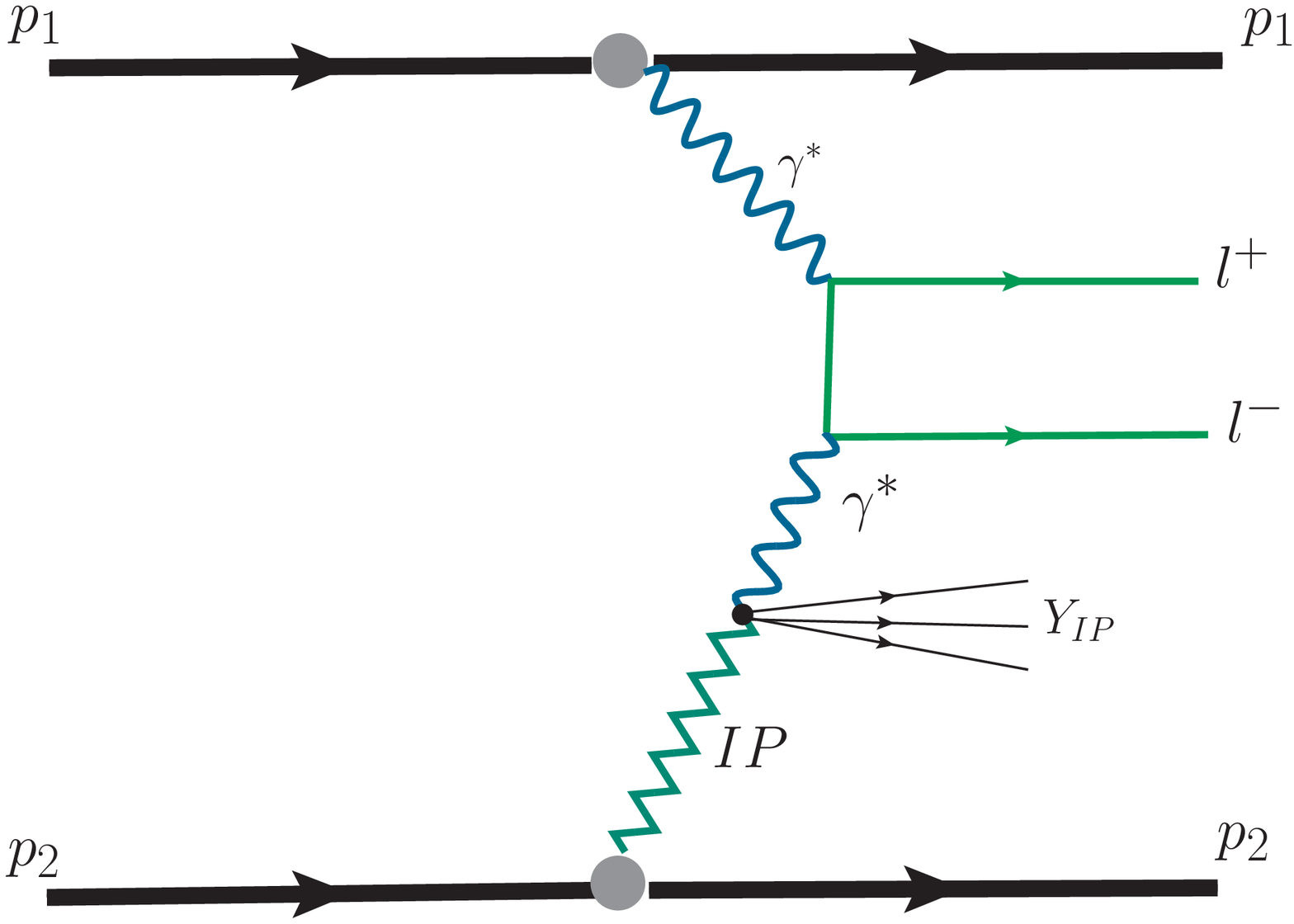}
\includegraphics[width=6cm]{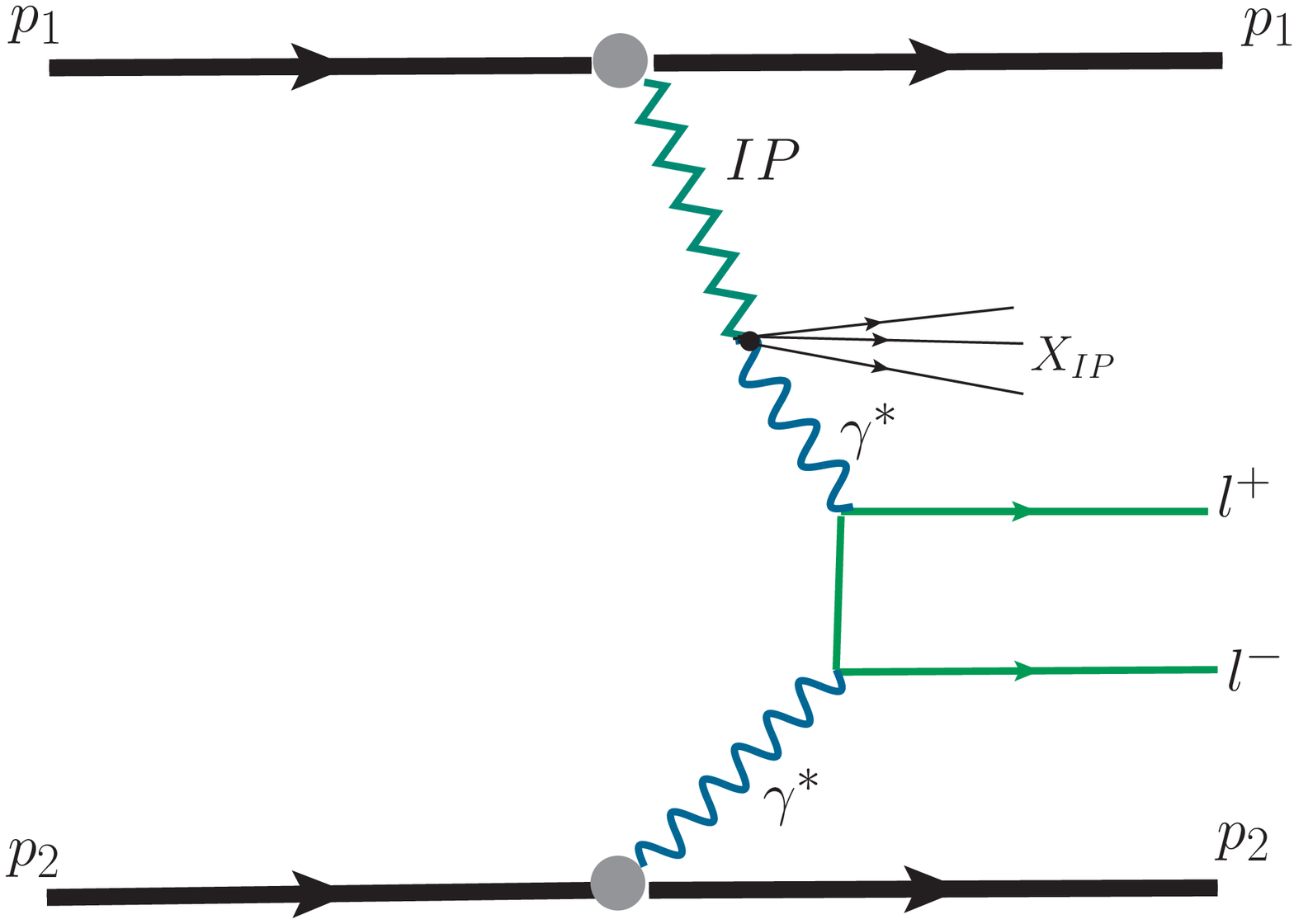}
\caption{Diffractive mechanisms of dilepton production
in proton-proton collisions.}
\label{fig:diffractive_diagrams}
\end{figure}

Only the diffractive mechanism shown in
Fig.\ref{fig:diffractive_diagrams} could lead to $\xi_i <$ 0.1.
However, the diffractive mechanism happens only in about 10 \% of all
cases as was measured at HERA \cite{RGE}. In addition, the pomeron
remnant would destroy the rapidity gap. Such a process was not discussed
in the context of $l^+ l^-$ production in $p p$ collisions with
rapidity gap requirement. Also the diffractive photon distribution
in pomeron was not discussed. One may expect:
\begin{equation}
\frac{d {\cal{F}}_{diff}}{d M^2}(x,q_t^2,M^2) \ll
\frac{d {\cal{F}}_{ine}}{d M^2}(x,q_t^2,M^2) \; .
\end{equation}
In addition, the pomeron remnant would destroy the rapidity gap and
the rapidity gap veto would almost totally eliminate contribution of 
such processes in the context of forward proton measurement
discussed in the present paper.

%-------------------------------------------------------------
%-------------------------------------------------------------

The discussion above shows that the double-dissociative contribution
would be to large extend eliminated.

The ATLAS collaboration analysis imposes the consistency requirements:
\begin{equation}
\xi_1 = \xi_{ll}^+  \; , \;  \xi_2 = \xi_{ll}^- \; .
\end{equation}
The longitudinal momentum fractions of the photons were calculated
in the ATLAS analysis as:
\begin{eqnarray}
\xi_{ll}^+ &=& \left( M_{ll}/\sqrt{s} \right) \exp(+Y_{ll}) \; , \nonumber \\
\xi_{ll}^- &=& \left( M_{ll}/\sqrt{s} \right) \exp(-Y_{ll}) \; .
\end{eqnarray}
Only lepton variables enter the formula.
This is only approximate formula which can be improved, if necessary.
We will use the same formula in our analysis.

%-----------------------
\section{Results}
%-----------------------

In the calculations described below we shall take typical cuts on dileptons:
-2.5 $< y_1, y_2 <$ 2.5 and $p_{1t}, p_{2t} >$ 15 GeV.
We shall show also results with extra cuts on $\xi_{ll}^+$ or
$\xi_{ll}^-$.
In the following we shall not exclude mass window around
$Z$-boson mass $m_Z$ and/or lepton acoplanarity, as was done in \cite{ATLAS}.

%-----------------------------------------------------
\subsection{Double-elastic contribution}
%-----------------------------------------------------

We start from presenting results for the double-elastic contribution.
Here the flux of photons can be expressed in terms of elastic
form factors of proton.
We take typical cuts for such a measurement \cite{ATLAS}:
$p_t >$ 15 GeV and -2.5 $< y_1, y_2 <$ 2.5.
In the following we limit ourself to dimuon production.
The results for electron production are very similar, at least for the
relatively high cuts on electron/muon transverse momenta.

%-------------------------------------------------------------
\begin{figure}
\includegraphics[width=7cm]{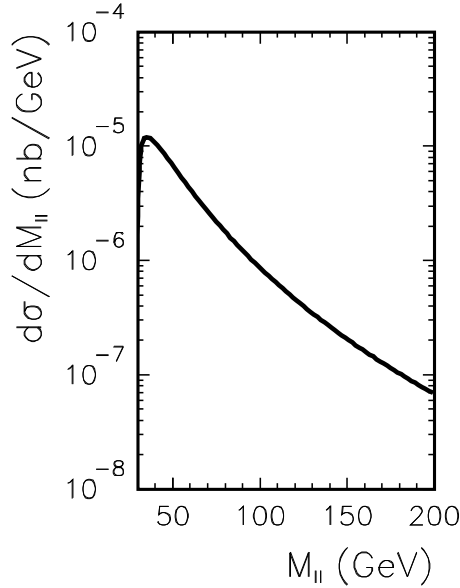}
\caption{Dimuon invariant mass distribution of the double-elastic
mechanism.}
\label{fig:dsig_dM}
\end{figure}
%-------------------------------------------------------------

\begin{figure}
\includegraphics[width=7cm]{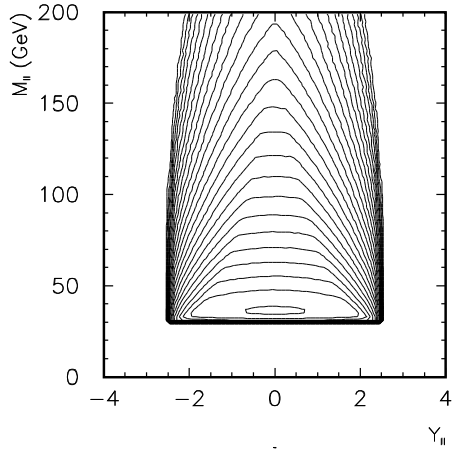}
\caption{Two-dimensional distribution in ($M_{ll},Y_{ll}$)
for double-elastic contribution. Here no cuts on neither $\xi_1$
nor $\xi_2$ were imposed. The $p_{t,\mu} >$ 15 GeV and 
-2.5 $< y_1, y_2 <$ 2.5 conditions were imposed here.
}
\label{fig:dsig_dMdY}
\end{figure}

Very interesting is the two-dimensional distribution in 
($M_{ll},Y_{ll}$) \cite{ATLAS}.
We show it for the double-elastic contribution in
Fig.\ref{fig:dsig_dMdY}.

%-------------------------------------------------------------
%-------------------------------------------------------------

\begin{figure}
\includegraphics[width=7cm]{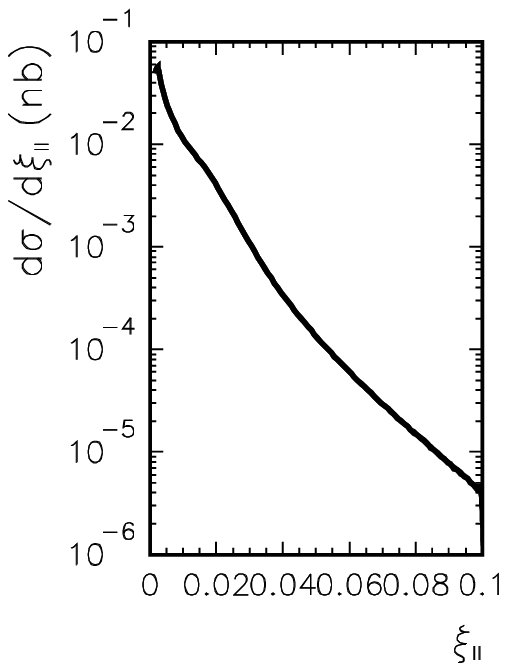}
\caption{$\xi_{1/2}$ distribution for dimuon production for 
the double-elastic mechanism.}
\label{fig:dsig_dxi_elaela}
\end{figure}

So far we did not include cuts on $\xi_1 = \xi_{ll}^+$ or 
$\xi_2 = \xi_{ll}^-$.
The distribution in $\xi_{ll}^{\pm}$ is shown in
Fig.\ref{fig:dsig_dxi_elaela}.
It is a steeply falling function with increasing $\xi$.\\

%{\bf We should understand why it finishes so early.}\\

%-------------------------------------------------------------
%-------------------------------------------------------------

\begin{figure}
\includegraphics[width=7cm]{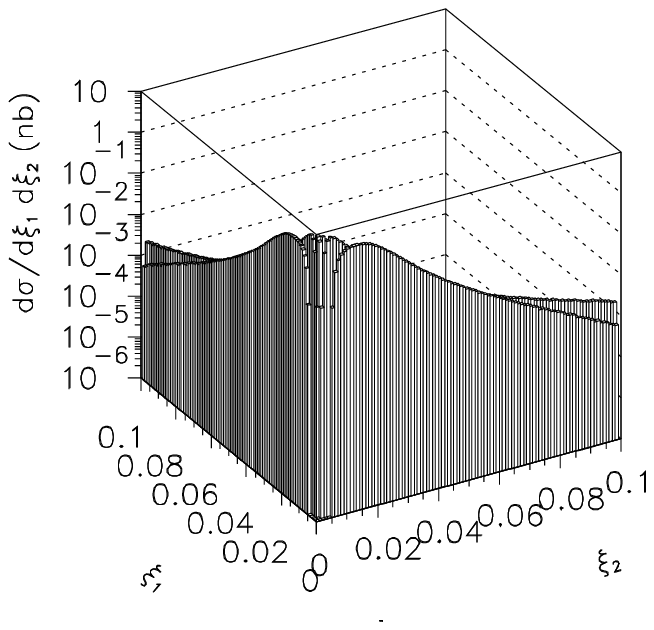}
\caption{Two-dimensional distribution in ($\xi_{ll}^+,\xi_{ll}^-$)
for the double-elastic mechanism.
}
\label{fig:dsig_dxi1dxi2}
\end{figure}

A two-dimensional distribution ($\xi_{ll}^+,\xi_{ll}^-$) 
is shown in Fig.\ref{fig:dsig_dxi1dxi2}.
A strong dependence on both $\xi_{ll}^+$ and $\xi_{ll}^-$ can
be observed.
This suggests that cuts on $\xi$'s will significantly lower
the measured cross section.

%-------------------------------------------------------------
%-------------------------------------------------------------

%In the recent study presented in \cite{ATLAS} only one photon was measured
%in AFP in order to get reasonable statistics. Typical experimental
%condition is:  $\xi_{min} < \xi_{i} < \xi_{max}$.
%For the ATLAS apparatus it is \cite{ATLAS}: 
%$\xi_{min} \approx$ 0.035, $\xi_{max} \approx$ 0.08.
%In the following we shall impose such a condition also in our calculations.
%We observe, see Fig.\ref{fig:dsig_dMdY_elaela_withcuts}, two separate bands 
%for conditions exclusively on $\xi_1$ (left panel)
%and for condition exclusively on $\xi_2$ (right panel). 
%Only for exteremely large $M_{ll} >$ 200 GeV one can satisfy both 
%conditions simultaneously.
%However, there the cross section is exteremely small.

%-------------------------------------------------------------
%\begin{figure}
%\includegraphics[width=7cm]{map_MY_elaela_cut1.eps}
%\includegraphics[width=7cm]{map_MY_elaela_cut2.eps}
%\caption{
%Two-dimension distribution in ($M_{ll},Y_{ll}$).
%for double-elastic contribution.
%Here we have imposed experimental condition on $\xi_1$ (left panel) or 
%$\xi_2$ (right panel) as explained in the main text.
%The $p_{t,\mu} >$ 15 GeV condition was imposed in addition.
%}
%\label{fig:dsig_dMdY_elaela_withcuts}
%\end{figure}
%-------------------------------------------------------------

%-------------------------------------------------------------
\subsection{Single-dissociative contribution}
%-------------------------------------------------------------

Now we wish to discuss similar distributions for single-dissociative
processes. 

As will be discussed below the transverse momenta of
initial (intermediate) photons in the inelastic vertex are large.
We integrate over $q_{2t} \in$ (0,100-500 GeV) for elastic-inelastic
and $q_{1t} \in$ (0,100-500 GeV) for inelastic-elastic contribution.\\
%{\bf The upper integration limit 50 GeV may be not sufficient,
%  especially for large $p_{t,sum}$.}

We start from the excitation of continuum.
In the calculation below, for illustration we use the Szczurek-Uleshchenko 
deep-inelastic structure function \cite{Szczurek:1999rd} 
in calculating inelastic photon flux.\\

\begin{figure}
\includegraphics[width=8cm]{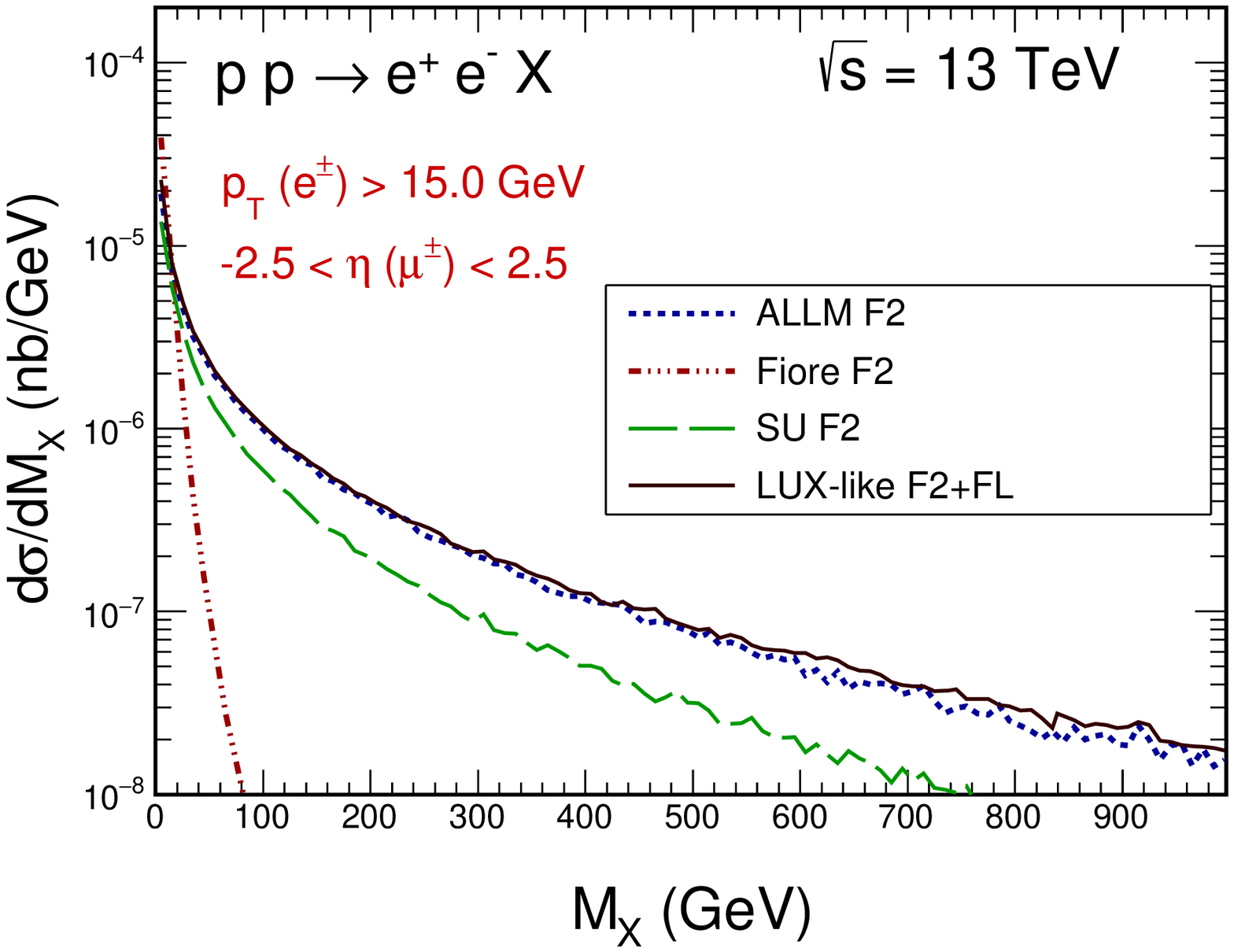}
\caption{Distribution in the mass of the baryonic remnant system ($M_X$ or
  $M_Y$) for different structure functions from the literature.
Here, in the case of the SU parametrization, only partonic contribution 
is included.}
\label{fig:dsig_dMX}
\end{figure}

In Fig.\ref{fig:dsig_dMX} we show the distribution in the mass of the
excited baryonic state. Here no extra cut on either $\xi_i$ or
$p_{t,pair}$ ($\vec{p}_{t,pair} = \vec{p}_1 + \vec{p}_2$) was imposed. 
The distribution for the ALLM and LUX-like structure functions 
are very similar and extend to very high
$M_X$ (or $M_Y$) masses. The Fiore at al. structure function
\cite{Fiore:2002re} does not lead to excitation of large
$M_X$ and/or $M_Y$ and is therefore reliable
only at low invariant masses.
The reason is a missing in this parametrization \cite{Fiore:2002re} 
partonic contribution.

%------------------------------------------------------------------
%------------------------------------------------------------------

%.......................................................\\
%.......................................................\\
%.......................................................\\

%In Fig.\ref{fig:dsig_dMdY_SD_withcuts} we show two-dimensional
%distribution in ($M_{ll},Y_{ll}$) for elastic-inelastic (left panel)
%and inelastic-elastic (right panel) single dissociative production
%of the dimuon pairs. We impose condition on $\xi_1$ for
%elastic-inelastic and on $\xi_2$ for inelastic-elastic contributions.\\
%{\bf We observe some asymmetry.}\\
%{\bf Symmetrization of the matrix element are needed.}

%-------------------------------------------------------------
%\begin{figure}
%\includegraphics[width=7cm]{map_MY_elaine_cut1.eps}
%\includegraphics[width=7cm]{map_MY_ineela_cut2.eps}
%\caption{
%Two-dimensional distribution in ($M_{ll},Y_{ll}$)
%for elastic-inelastic (left panel) and inelastic-elastic (right panel) 
%contributions.
%The Szczurek-Uleshchenko structure function parametrization was used
%here for illustration.
%Here we have imposed experimental condition on $\xi_1$ (left panel) or 
%$\xi_2$ (right panel) as explained in the main text.
%The $p_{t,\mu} >$ 15 GeV condition has been imposed in addition.
%}
%\label{fig:dsig_dMdY_SD_withcuts}
%\end{figure}
%-------------------------------------------------------------

\begin{figure}
\includegraphics[width=7cm]{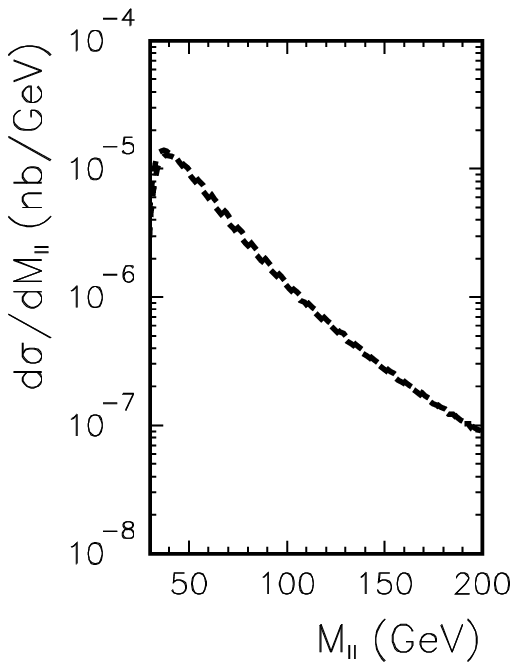}
\caption{Distribution in dilepton invariant mass for
elastic-inelastic and inelastic-elastic contributions.
Here the SU structure function parametrization was used for example.
Here the cuts on $\xi_{ll}^{+}$ or $\xi_{ll}^-$ are not imposed.}
\label{fig:dsig_dMll_noxicuts}
\end{figure}

Now let us show some more differential distributions.
In Fig.\ref{fig:dsig_dMll_noxicuts} we show dilepton invariant mass
distribution without any cuts on $\xi_{ll}^+$ or $\xi_{ll}^-$.
%There is some small disagreement between the elastic-inelastic
%and inelastic-elastic contribution. This seems to be represent 
%accuracy of the calculation.

%-------------------------------------------------------------
%-------------------------------------------------------------

%The distributions in $Y_{ll}$ are shown in
%Fig.\ref{fig:dsig_dYll_noxicuts}. Here we show only elastic-inelastic
%and inelastic-elastic contributions. In this case the distributions
%are not identical, but must be symmetric with respect to $Y_{ll}$ = 0.\\
%{\bf There seems to be a small asymmetry ?}

%--------------------------------------------------------------
%\begin{figure}
%\includegraphics[width=7cm]{dsig_dYll_xillcuts.eps}
%\caption{Distribution in dilepton rapidity for
%four different contributions considered.
%The solid lines represent contributions of fully elastic and the dashed
%line represent contributions of single-dissociative proceseses.
%Here the cuts on $\xi_{ll}^{+}$ or $\xi_{ll}^{-}$ are imposed.}
%\label{fig:dsig_dYll_noxicuts}
%\end{figure}
%--------------------------------------------------------------

\begin{figure}
\includegraphics[width=7cm]{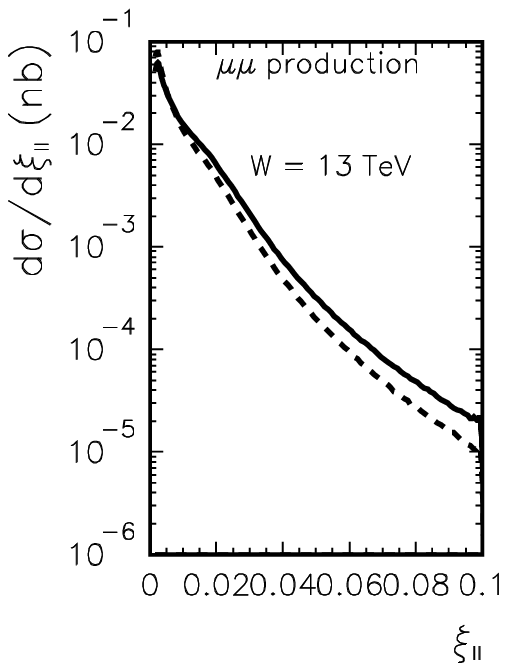}
\caption{Distribution in $\xi_{ll}^{\pm}$ for elastic-inelastic and
inelastic-elastic contributions. In this calculation the ALLM
parametrization was used.
The solid line is for right cuts on proton from the elastic vertex 
($\xi_{ll}^+$ for elastic-inelastic and $\xi_{ll}^-$ for
inelastic-elastic contributions),
the dashed line is for ``incorrect'' cuts on the inelastic system.
}
\label{fig:dsig_dxill_noxicuts}
\end{figure}

Finally in Fig.\ref{fig:dsig_dxill_noxicuts} we show distributions
in $\xi_{ll}^+$ or $\xi_{ll}^-$ for elastic-inelastic (black solid) 
and inelastic-elastic (black solid) contributions.
For completeness we show also (dashed line) distributions in $\xi_{ll}$
corresponding to the remnant systems.
The two distributions are fairly similar.

%-----------------------------------------------------------------
%-----------------------------------------------------------------

\begin{figure}
\includegraphics[width=7cm]{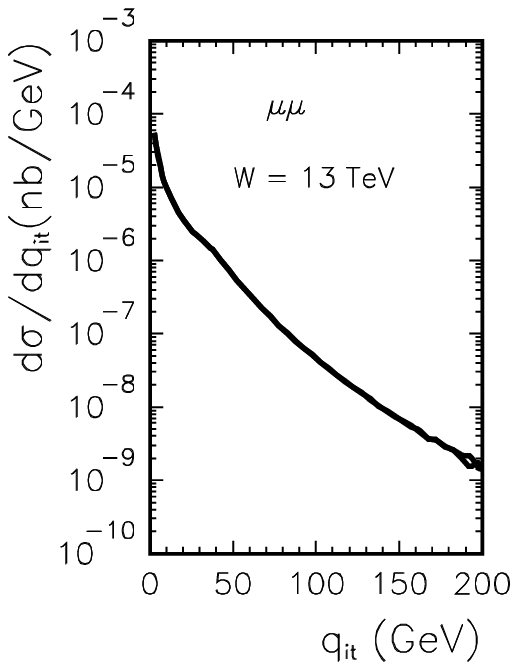}
\caption{Distribution in $q_{1/2,t}$ in the inelastic vertex.
Here the ALLM parametrization is used.}
\label{fig:dsig_dqit_noxicuts}
\end{figure}

In Fig.\ref{fig:dsig_dqit_noxicuts} we show distribution in the
initial photon transverse momentum in the inelastic vertex
for elastic-inelastic and inelastic-elastic contributions.
The distributions extend to large transverse momenta.

%-----------------------------------------------------------------
%------------------------------------------------------------------

\begin{figure}
\includegraphics[width=7cm]{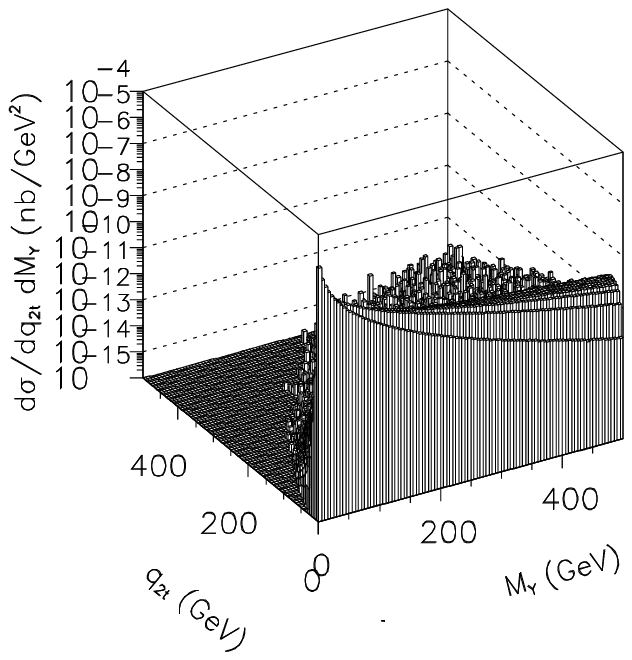}
\includegraphics[width=7cm]{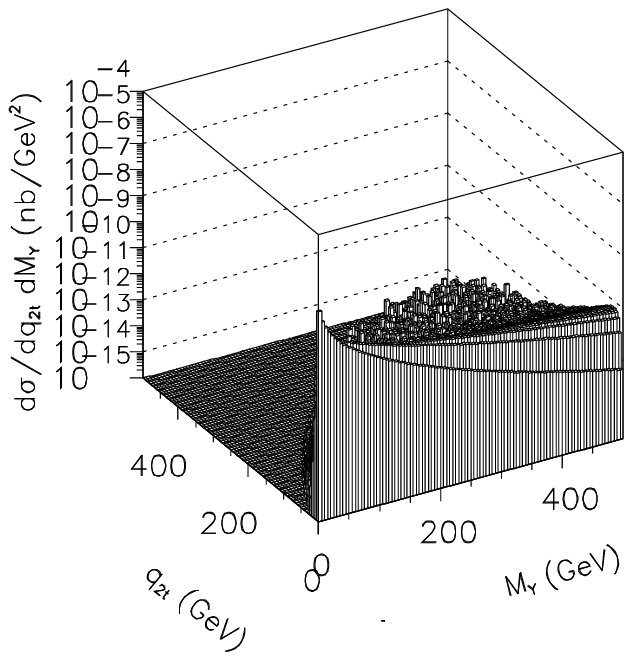}
\caption{Two-dimensional distribution in $(q_{2t},M_Y)$ 
for elastic-inelastic contribution. Similar distribution in
$(q_{1t},M_X)$ plane can be obtained for the inelastic-elastic
contribution.
We show results without $\xi$ cut (left panel) and with $\xi$ cut (right
panel).
}
\label{fig:dsig_dq2tdMY}
\end{figure}

Finally in this subsection we wish to discuss effect of correlations
between $q_{2t}$ and $M_Y$ for elastic-inelastic
and $q_{1t}$ and $M_X$ for inelastic-elastic contributions.
In Fig.\ref{fig:dsig_dq2tdMY} we show only the first case.
For small masses of the remnant only small $q_{2t}$ are generated.
In general, the larger $M_Y$ the larger $q_{2t}$ can be generated.
This shows that in the VEGAS integration one should carefully
adjust the limits of integration on the $(q_{2t},M_Y)$ plane.
There is no significant effect of the cut on the two-dimensional
distribution. 

%------------------------------------------------------------------
%-------------------------------------------------------------------

%----------------------------------------------------
\subsection{$\xi_{ll}^+$ or $\xi_{ll}^-$ cuts}
%----------------------------------------------------

In the recent study presented in \cite{ATLAS} only one photon was measured
in AFP in order to get reasonable statistics. Typical experimental
condition is:  $\xi_{min} < \xi_{i} < \xi_{max}$.
For the ATLAS apparatus it is \cite{ATLAS}: 
$\xi_{min} \approx$ 0.035, $\xi_{max} \approx$ 0.08.
In the following we shall impose such a condition also in our calculations.
Here we show how the $\xi_{ll}^+$ and $\xi_{ll}^-$ cuts work in
practice. 

%For illustration we take 0.035 $< \xi_{ll}^{\pm} < $ 0.08 
%from the recent ATLAS paper \cite{ATLAS}.

\begin{figure}
\includegraphics[width=6cm]{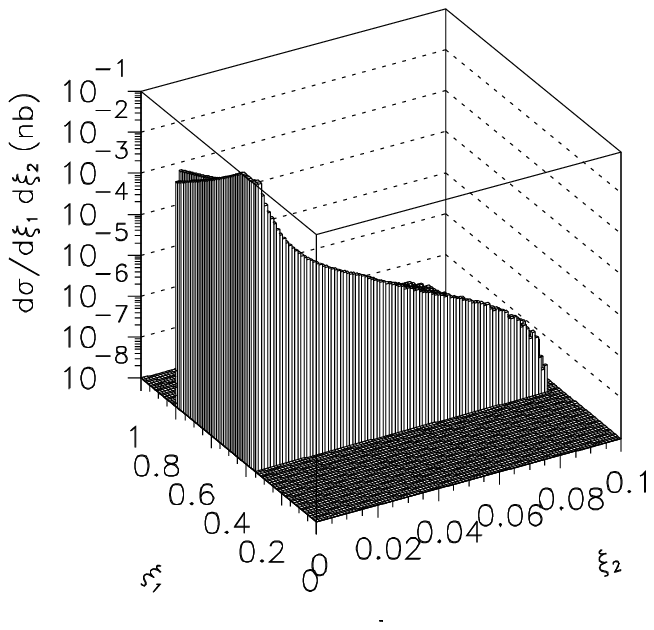}
\includegraphics[width=6cm]{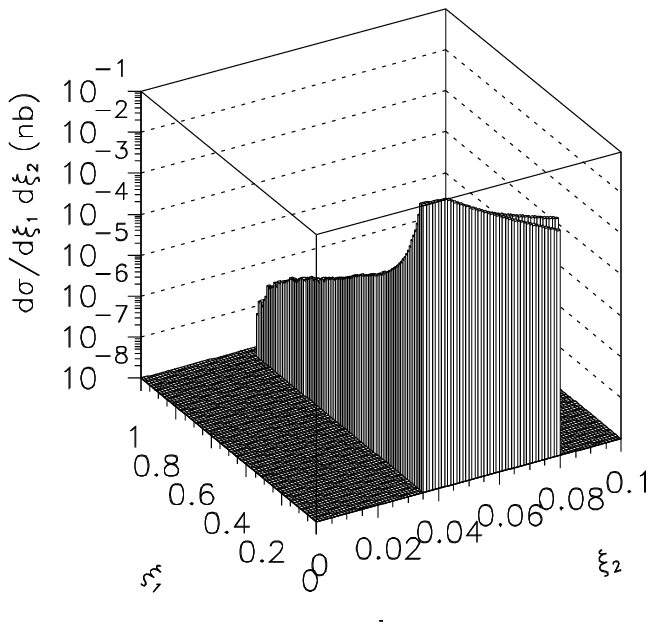}\\
\includegraphics[width=6cm]{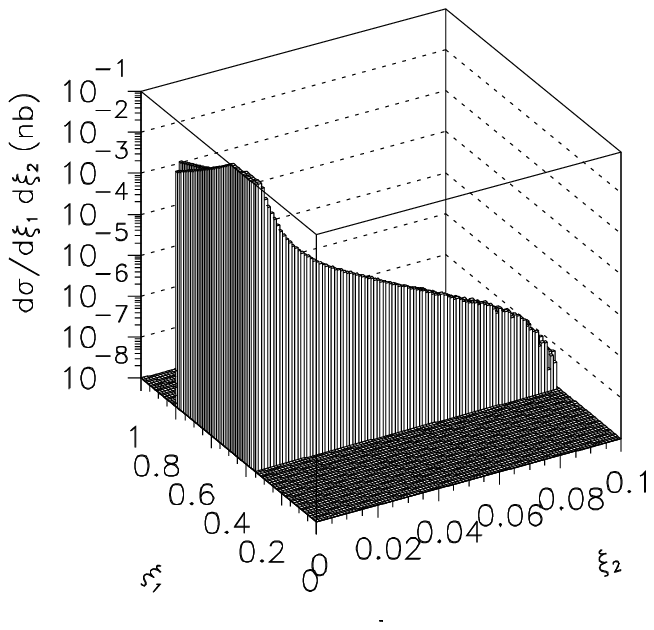}
\includegraphics[width=6cm]{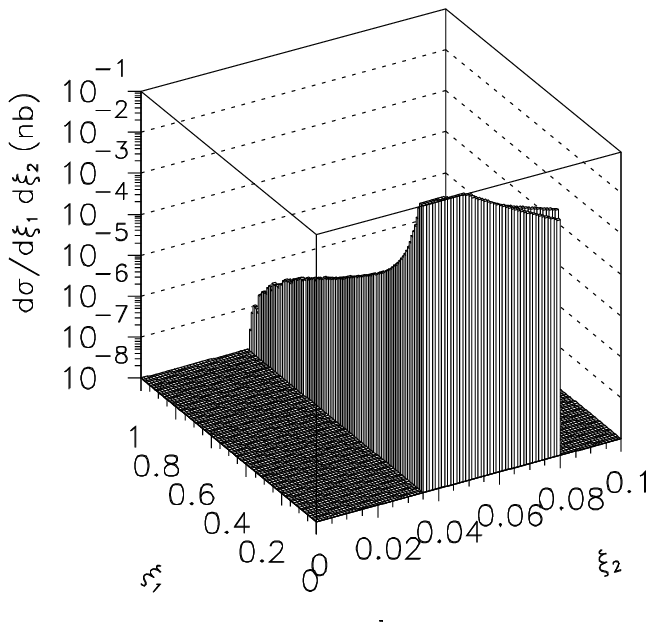}
\caption{
Two-dimensional distributions in ($\xi_{ll}^+,\xi_{ll}^-$) for
elastic-elastic, cut on 1 (upper-left corner),
elastic-elastic, cut on 2 (upper-right corner),
elastic-inelastic, cut on 1 (lower-left corner) and
inelastic -elastic, cut on 2 (lower-right corner).
Here the SU structure function was used for illustration.
}
\label{fig:map_xillpxillm}
\end{figure}

In Fig.\ref{fig:map_xillpxillm} we present two-dimensional distributions
in  $\xi_{ll}^+$ and $\xi_{ll}^-$ with appropriate cuts imposed
to illustrate the large reduction of the cross section.
%{\bf Our cuts work OK.}

%-----------------------------------------------------------------
%-----------------------------------------------------------------

The ATLAS experimental cuts make the integrated cross section rather low
compared to the result without such cuts as is shown in 
Table \ref{table:cs_with_xicuts}.\\
%{\bf The asymmetry of elastic-inelastic and inelastic-elastic
%  contribution is probably of numerical character ?}
The contribution of elastic-inelastic and inelastic-elastic mechanisms
seem bigger than that of the elastic-elastic one.
However, the situation changes when imposing extra cut on $p_{t,pair}$
(the numbers in paranthesis for single dissociative contributions).
We present also naive (no $\Omega$ factor, see 
\cite{Szczurek:1999rd}.
%\cite{SU..}) VDM contribution
We have calculated also contribution of resonance excitations
as parametrized in \cite{Fiore:2002re}.
The resonance contribution ($\Delta^+$ etc.) is about 15 \% of 
the elastic-elastic contribution.
The $\Delta^+$ contribution is the dominant resonance contribution.
We also use a popular in the context of HERA physics structure function 
parametrization of Abramowicz et al. \cite{Abramowicz:1997ms}.
We also show a result of a slightly modified SU parametrization.
The new Szczurek parametrization includes partonic and VDM contributions
as in the original Szczurek-Uleshchenko parametrization and includes 
in addition resonance contributions as parametrized by Fiore 
et al.\cite{Fiore:2002re}.
The LUX-like structure function (see \cite{LUX}) leads to largest 
cross section, especially when the cut on $\xi$-variable is imposed.\\
%{\bf We do not understand these huge numbers in the latter case ???}
These results were obtained with the generator (GEN) version of our
code.
In the case of LUX-like model we included also longitudinal
structure function $F_L$, which lowers the cross section
(see the numbers in the table).

%---------------------------------------------------------------------
\begin{table}
   \centering
   \caption{Integrated cross section for $\mu^+ \mu^-$ production in fb
   with one proton in the  0.035 $ < \xi_{ll}^{\pm} < $ 0.08 interval.
   In this calculation $p_{1t}, p_{2t} >$ 15 GeV and
   -2.5 $< y_1, y_2 <$ 2.5. No gap survival factor was imposed here.
   In the paranthesis we show result with extra $p_{t,pair} <$ 5 GeV
   condition. 2UN means the version of our code with doubly unintegrated
   photon distribution and GEN generator version of our code. 
   In other cases singly unintegrated photon
   distribution is used from a simplified version of our code.
   The numbers in [...] were obtained with exact formula for $\xi_1$ and
   $\xi_2$.
}
\begin{tabular}{|c|c|c|}
\hline
contribution & c.s. in fb without $\xi$-cuts & c.s. in fb with $\xi$-cuts  \\
\hline
elastic-elastic, cut on proton 1    & 358.68 &  5.4591  \\
elastic-elastic, cut on proton 2    & ...... &  5.4592  \\
\hline
%elastic-inelastic, cut on proton 1, SU, 0-50 GeV & 405.00 &  9.0178 (3.3490) \\
%inelastic-elastic, cut on proton 2  SU, 0-50 GeV & 405.92 &  9.0139 (3.3493) \\
%elastic-inelastic, cut on proton 1, SU, 0-100 GeV & 427.8949 & 10.0190 (3.3492) \\
%inelastic-elastic, cut on proton 2  SU, 0-100 GeV & 427.0130 & 10.0186 (3.3491) \\
%\hline
elastic-inelastic, VDM (no $\Omega$), 0-100 GeV & 98.0215 (2UN) &      \\
inelastic-elastic, VDM (no $\Omega$), 0-100 GeV & 98.0297 (2UN) &      \\
elastic-inelastic                SU partonic    & 449.1076 (2UN) &      \\ 
inelastic-elastic                SU partonic    & 449.0985 (2UN) &      \\
elastic-inelastic, cut on proton 1, ALLM  & 468.6102 (2UN) &  11.8292 \\
inelastic-elastic, cut on proton 2, ALLM  & 468.6102 (2UN) &  11.8294 \\
elastic-inelastic, new Szczurek           & 461.5330 (2UN) &  12.6046 [14.1806] (5.9311) \\
inelastic-elastic, new Szczurek           & 461.5750 (2UN) &  12.6032 [14.1806] (5.9309) \\
elastic-inelastic, new Szczurek, $M_Y >$ 500 GeV  &  ....  &   0.7152 \\
inelastic-elastic, new Szczurek, $M_X >$ 500 GeV  &  ....  &   0.7149 \\
\hline
elastic-inelastic, ALLM                   & 571.871 (GEN)  &  9.711 \\ 
inelastic-elastic, ALLM                   & 571.562 (GEN)  &  9.621 \\
elastic-inelastic, LUX-like, $F_2+F_L$    & 635.215 (GEN)  &  19.894 \\ 
inelastic-elastic, LUX-like, $F_2+F_L$    & 635.102 (GEN)  &  19.831 \\
elastic-inelastic, LUX-like, $F_2$ only   & ....... (GEN)  &  ...... \\ 
inelastic-elastic, LUX-like, $F_2$ only   & 656.702 (GEN)  &  ...... \\
\hline
elastic-inelastic, cut on proton 1, resonances  & 38.6709 (2UN) & 0.57872 \\
inelastic-elastic, cut on proton 2  resonances  & 38.6639 (2UN) & 0.57872 \\
elastic-inelastic, cut on proton 1, $\Delta^+$  & 28.5844 (2UN) & 0.42755 \\
inelastic-elastic, cut on proton 2  $\Delta^+$  & 28.5814 (2UN) & 0.42763 \\
\hline
\end{tabular}
\label{table:cs_with_xicuts}
\end{table}
%---------------------------------------------------------------------

A technical remark is in order here.
The range of integration in $q_{1t}$ and $q_{2t}$ is crucial to get
correct result. The limits of integration should be different for
elastic and inelastic arm.
For double-elastic contribution it is sufficient to take
$q_{i,t} <$ 5 GeV as an upper integration limit.
For single dissociative contribution we integrate in the interval
$q_{i,t} <$ 100-500 GeV or $q_{i,t} <$ 100-500 GeV for the inelastic arm. 
The first limit is enough when $p_{t,pair} <$ 5 GeV is imposed
as in the ATLAS experiment, otherwise it should be a larger limit.

\begin{figure}
\includegraphics[width=7cm]{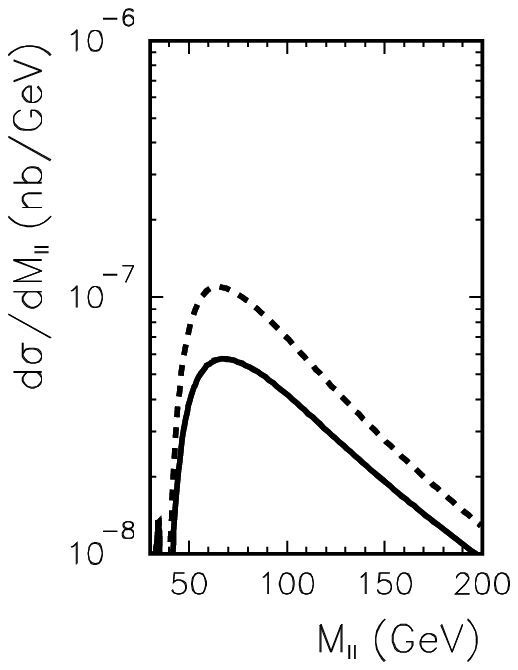}
\caption{Distribution in dilepton invariant mass for
the different contributions considered.
Here the cuts on $\xi_{ll}^{+}$ or $\xi_{ll}^-$ are imposed.
The solid line is for double elastic contribution and the dashed line is
for single dissociation contribution.}
\label{fig:dsig_dMll_withcuts}
\end{figure}

Now we wish to show other differential distributions for the case
of including the experimental cuts on $\xi_{ll}^{\pm}$.
In Fig.\ref{fig:dsig_dMll_withcuts} we show dilepton invariant mass
distribution. The two coinciding solid lines correspond to
elastic-elastic contributions, while the two dashed lines to
single dissociative contributions. \\
%{\bf They do not coincide as they should.} \\

%-------------------------------------------------------------
%-------------------------------------------------------------

The correlation in $(Y_{ll}, M_{ll})$ becomes very interesting when
cuts on $\xi$ are imposed.

\begin{figure}
\includegraphics[width=7cm]{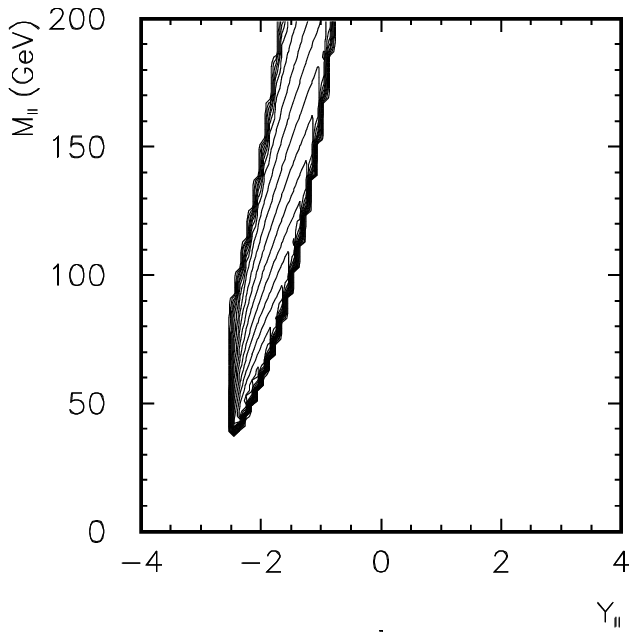}
\includegraphics[width=7cm]{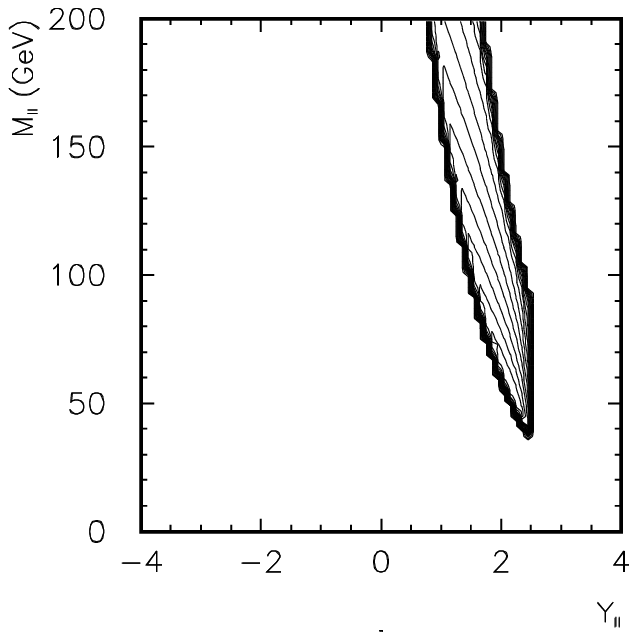}
\caption{
Two-dimension distribution in ($M_{ll},Y_{ll}$)
for double-elastic contribution.
Here we have imposed experimental condition on $\xi_2$ (left panel) or 
$\xi_1$ (right panel) as explained in the main text.
The $p_{t,\mu} >$ 15 GeV condition was imposed in addition.
}
\label{fig:dsig_dMdY_elaela_withcuts}
\end{figure}

Let us start from elastic-elastic contribution.
We observe, see Fig.\ref{fig:dsig_dMdY_elaela_withcuts}, two separate bands 
for conditions exclusively on $\xi_1$ (left panel)
and for condition exclusively on $\xi_2$ (right panel). 
Only for exteremely large $M_{ll} >$ 200 GeV one can satisfy both 
conditions simultaneously.
However, there the cross section is exteremely small.

%-------------------------------------------------------------
%-------------------------------------------------------------

\begin{figure}
\includegraphics[width=7cm]{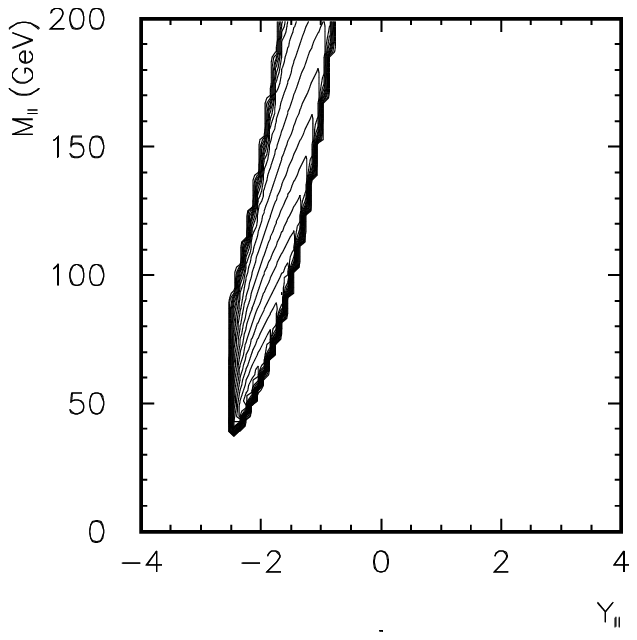}
\includegraphics[width=7cm]{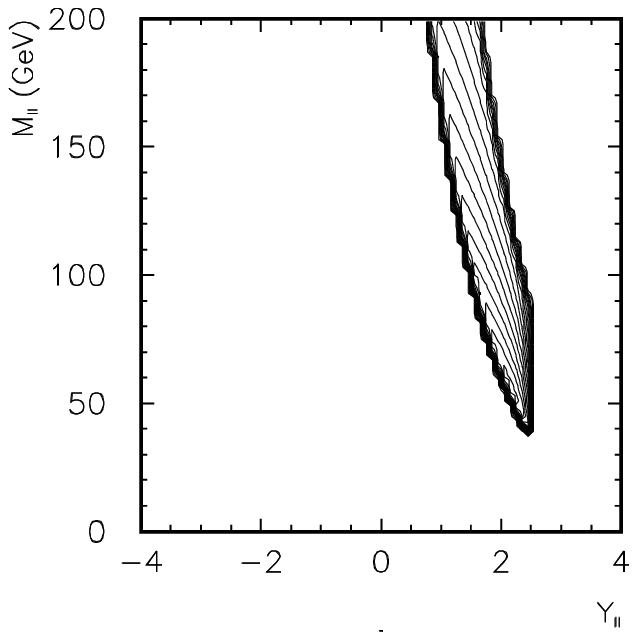}
\caption{
Two-dimensional distribution in ($M_{ll},Y_{ll}$)
for inelastic-elastic (left panel) and elastic-inelastic (right panel) 
contributions.
The Szczurek-Uleshchenko structure function parametrization was used
here for illustration.
Here we have imposed experimental condition on $\xi_2$ (left panel) or 
$\xi_1$ (right panel) as explained in the main text.
The $p_{t,\mu} >$ 15 GeV condition has been imposed in addition.
}
\label{fig:dsig_dMdY_SD_withcuts}
\end{figure}

In Fig.\ref{fig:dsig_dMdY_SD_withcuts} we show two-dimensional
distribution in ($M_{ll},Y_{ll}$) for elastic-inelastic (left panel)
and inelastic-elastic (right panel) single dissociative production
of the dimuon pairs. We impose condition on $\xi_1$ for
elastic-inelastic and on $\xi_2$ for inelastic-elastic contributions.\\
%{\bf We observe some asymmetry.}\\
%{\bf Symmetrization of the matrix element are needed.}

%-------------------------------------------------------------
%-------------------------------------------------------------

\begin{figure}
\includegraphics[width=7cm]{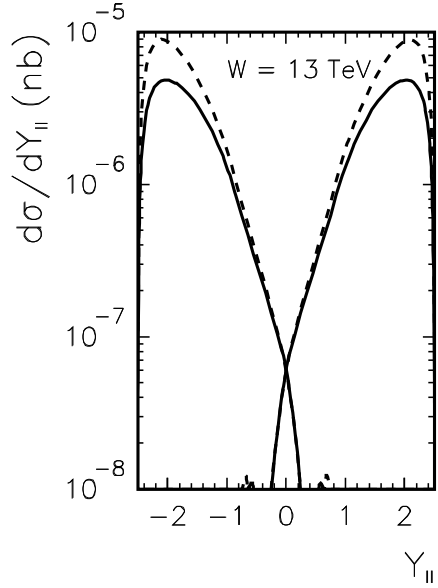}
\caption{Distribution in dilepton rapidity for
four different contributions considered.
Here the cuts on $\xi_{ll}^{+}$ or $\xi_{ll}^-$ are imposed.
The solid line is for double elastic contribution and the dashed line is
for single dissociation contribution.}
\label{fig:dsig_dYll_withcuts}
\end{figure}

Let us look now at the projection on $Y_{ll}$.
The distribution in rapidity of the pair $Y_{ll}$ is shown in 
Fig.\ref{fig:dsig_dYll_withcuts}.
We observe clear symmetry with respect to $Y_{ll}$ = 0 for the
two elastic-elastic contributions and similar symmetry between
elastic-inelastic and inelastic-elastic contributions.
The contribution related to the cut on $\xi_{ll}^+$ or $\xi_{ll}^-$
are almost totally separated. This is a reason of a dip observed
in $d \sigma / d Y_{ll}$ at $Y_{ll}$ = 0 by 
the ATLAS collaboration \cite{ATLAS}.

%--------------------------------------------------------------
%--------------------------------------------------------------

\begin{figure}
\includegraphics[width=7cm]{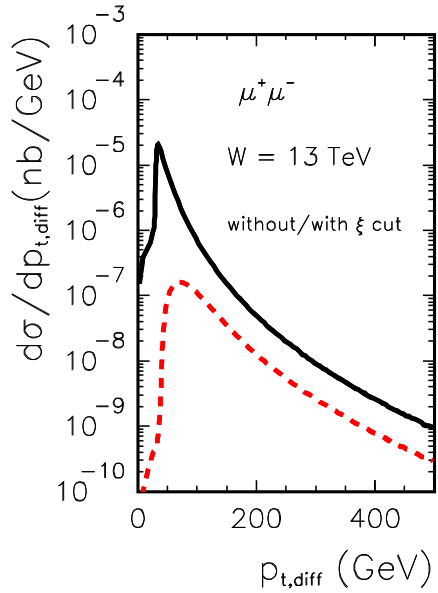}
\caption{Distribution in $p_{t,diff}$ without (upper solid coinciding curves)
and with (lower dashed coinciding curves) $\xi$ cuts.}
\label{fig:dsig_dptdiff}
\end{figure}

A final fully leptonic variable we wish to consider in the context
of the $\xi$ cut is $p_{t,diff}$, where 
$\vec{p}_{t,diff} = \vec{p}_{1t} - \vec{p}_{2t}$.
We show corresponding distributions in Fig.\ref{fig:dsig_dptdiff}.
The biggest effect of the cut is for small values of $p_{t,diff}$. 

%----------------------------------------------------------------------
%----------------------------------------------------------------------

\begin{figure}
\includegraphics[width=5.5cm]{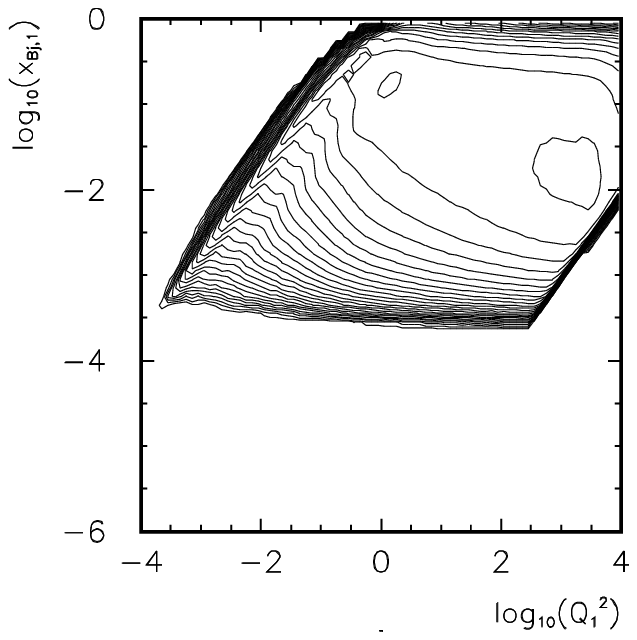}
\includegraphics[width=5.5cm]{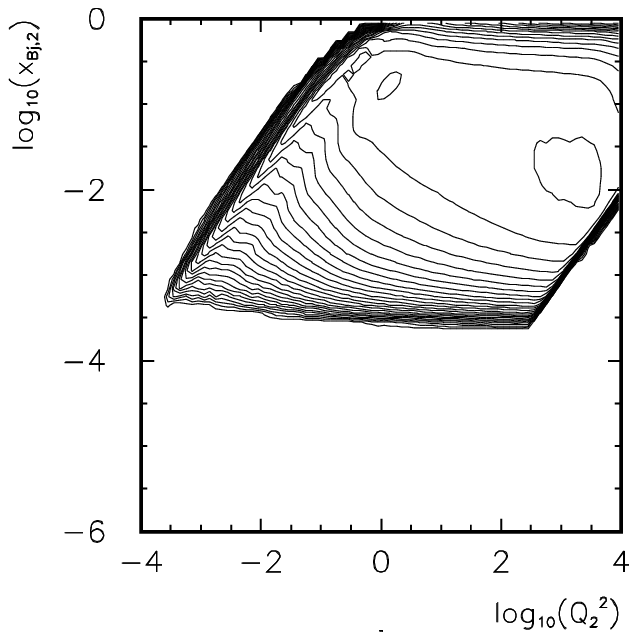}
\includegraphics[width=5.5cm]{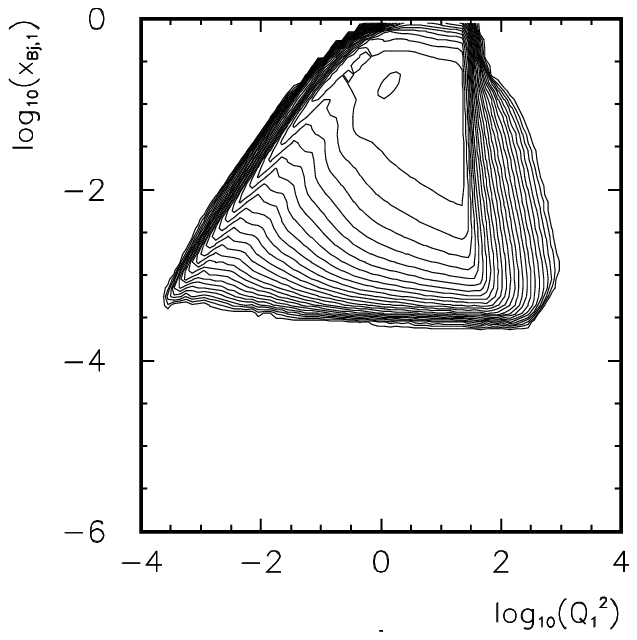}
\includegraphics[width=5.5cm]{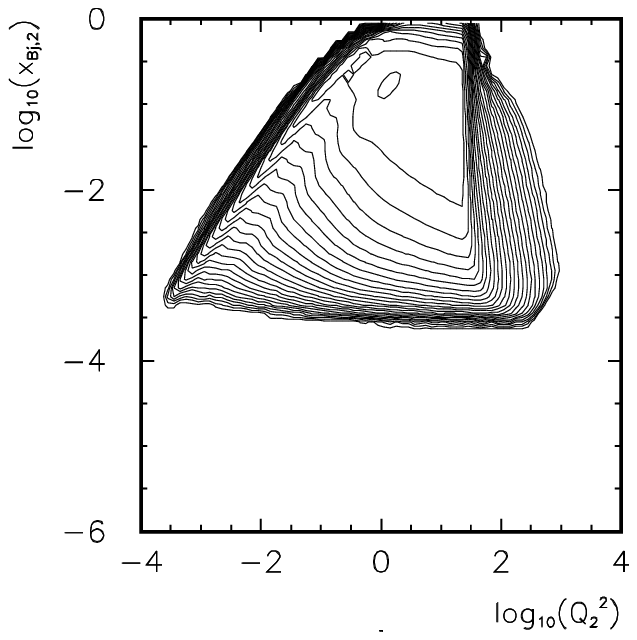}
\caption{The range in the ($x_{Bj},Q^2$) space tested in
inelastic-elastic (left) and elastic-inelastic DIS processes
with $\xi$ cuts. The lower panels include also an extra cut
$p_{t,pair} <$ 5 GeV.}
\label{fig:xBj-Q2}
\end{figure}

What are typical $x_{Bj}$ and $Q^2$, i.e. arguments of the structure
functions for the considered processes with single dissociation
(inelastic-elastic or elastic-inelastic) is shown in Fig.\ref{fig:xBj-Q2}.
Both perturbative ($Q^2 >$ 2 GeV$^2$) and nonperturbative 
($Q^2 <$ 2 GeV$^2$) regions enter the corresponding calculations.
The nonperturbative region is even relatively larger
when the cut on $p_{t,pair} <$ 5 GeV is imposed as in the recent
ATLAS \cite{ATLAS} paper.

%....................................................\\
%.....................................................\\

%------------------------------------------------------------------
%-------------------------------------------------------------------

\begin{figure}
\includegraphics[width=7cm]{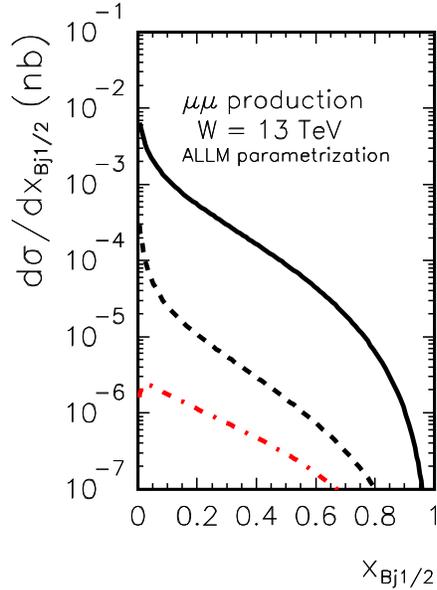}
\caption{Distribution in $x_{Bj}$ for single dissociative process.
Shown are results without (solid line) and with (dashed line)
cuts on longitudinal momentum fraction $\xi$. In this calculation 
the ALLM parametrization of $F_2$ structure function is used.
For completeness we show also contribution of proton resonances with 
cuts on $\xi$ (red dash-dotted line).}
\label{fig:dsig_dxBj}
\end{figure}

In Fig.\ref{fig:dsig_dxBj} we show distribution (projection of the
previous two-dimensional distribution) in $x_{Bj}$ for 
the ALLM parametrization without (solid line) and with (dashed line)
cuts on $\xi$. The two distributions have rather similar shape
which means that a similar range of $x_{Bj}$ is tested in both
cases. Here the most probable range of $x_{Bj}$ is about 10$^{-2}$,
the region of experiments on deep-inelastic scattering performed
in the past by the NMC collaboration. In the $\gamma \gamma \to l^+ l^-$
process one is not sensitive to a small $x_{Bj}$ region ($x_{Bj} <$ 10$^{-3}$).
For comparison we also show distribution for the resonant contributions
that are not explicit in most of the parametrizations used
in the literature.

%---------------------------------------------------------------
%---------------------------------------------------------------

\begin{figure}
\includegraphics[width=7cm]{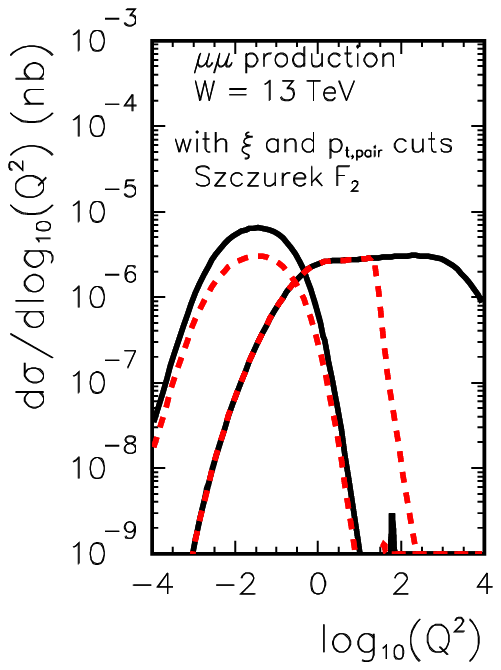}
\caption{Distribution in $log_{10}(Q_i^2)$ for single dissociative
  process with the cut on $\xi$.  We show distributions for 
  elastic (left) and inelastic (right) vertex.
  In this calculation the new Szczurek parametrization of $F_2$ was used. 
  We also show similar distributions with the upper cut on 
  $p_{t,pair}$ (red dashed line) as applied 
  in the recent ATLAS measurement \cite{ATLAS}.
}
\label{fig:dsig_dxiQ2}
\end{figure}

In Fig.\ref{fig:dsig_dxiQ2} we show complementary distribution
in the second argument of the structure function ($Q_1^2$ or $Q_2^2$).
Both small (nonperturbative) and large (perturbative) $Q_i^2$ occur.
The nonperturbative region is relatively larger when the experimental
cut on $p_{t,pair} <$ 5 GeV \cite{ATLAS} is imposed.

%----------------------------------------------------------------
%----------------------------------------------------------------

\begin{figure}
\includegraphics[width=7cm]{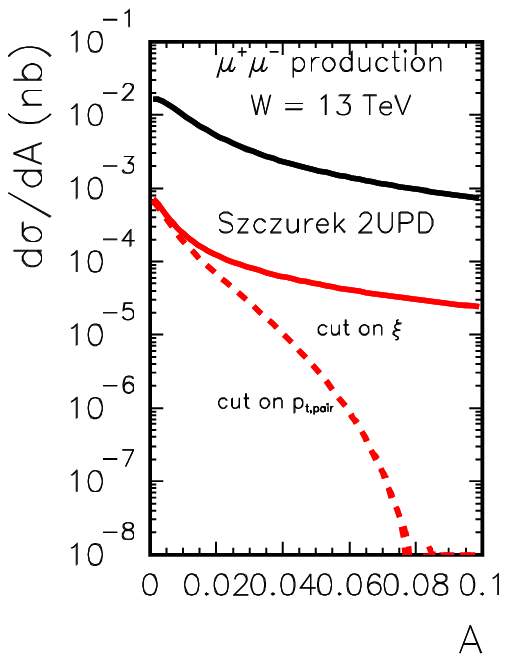}
\caption{Acoplanarity distribution for single dissociative contributions
without any (upper black solid curve), with $\xi$ cut 
(middle red solid curve) and with
extra $p_{t,pair} <$ 5 GeV condition (lower red dashed curve).
In this calculation the new Szczurek parametrization of $F_2$ 
(including resonances) was used. 
}
\label{fig:dsig_dacop}
\end{figure}

How the acoplanarity distribution depends on cuts is illustrated
in Fig.\ref{fig:dsig_dacop}.
We show result without any cut, with $\xi$ cut and with additional
condition on the pair transverse momentum.
The condition on the pair transverse momentum significantly change
the distribution. Also photon final state radiation may be important
in this context but this goes beyond the scope of the present analysis.

%-----------------------------------------------------------------
%--------------------------------------------------------------------

\begin{figure}
\includegraphics[width=6.5cm]{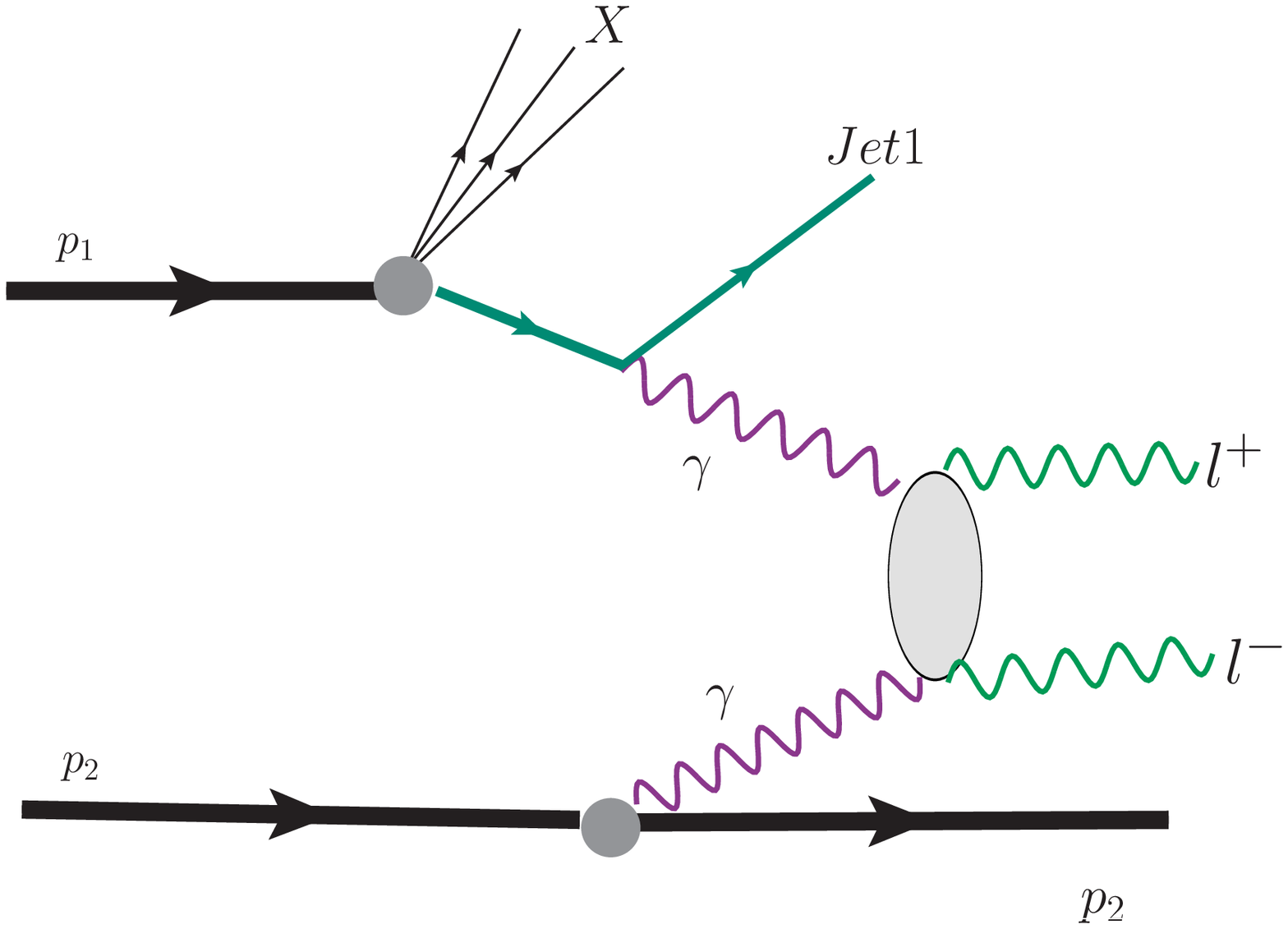}
\caption{Diagram with (min)jet production due to quark/antiquark knock-out.}
\label{fig:jet-diagram}
\end{figure}

The single dissociative process leads to an emission of a (mini)jet
(see Fig.\ref{fig:jet-diagram})
which produces hadrons (mostly pions) that may destroy the rapidity 
gap if required experimentally. Here we wish to show rapidity
distributions of such jets separately for elastic-inelastic and 
inelastic-elastic contributions. We show results without and with
cut on $\xi_{1/2}$.

%----------------------------------------------------------------------
%----------------------------------------------------------------------

%------------------------------------------------------------------
\begin{figure}
\includegraphics[width=8cm]{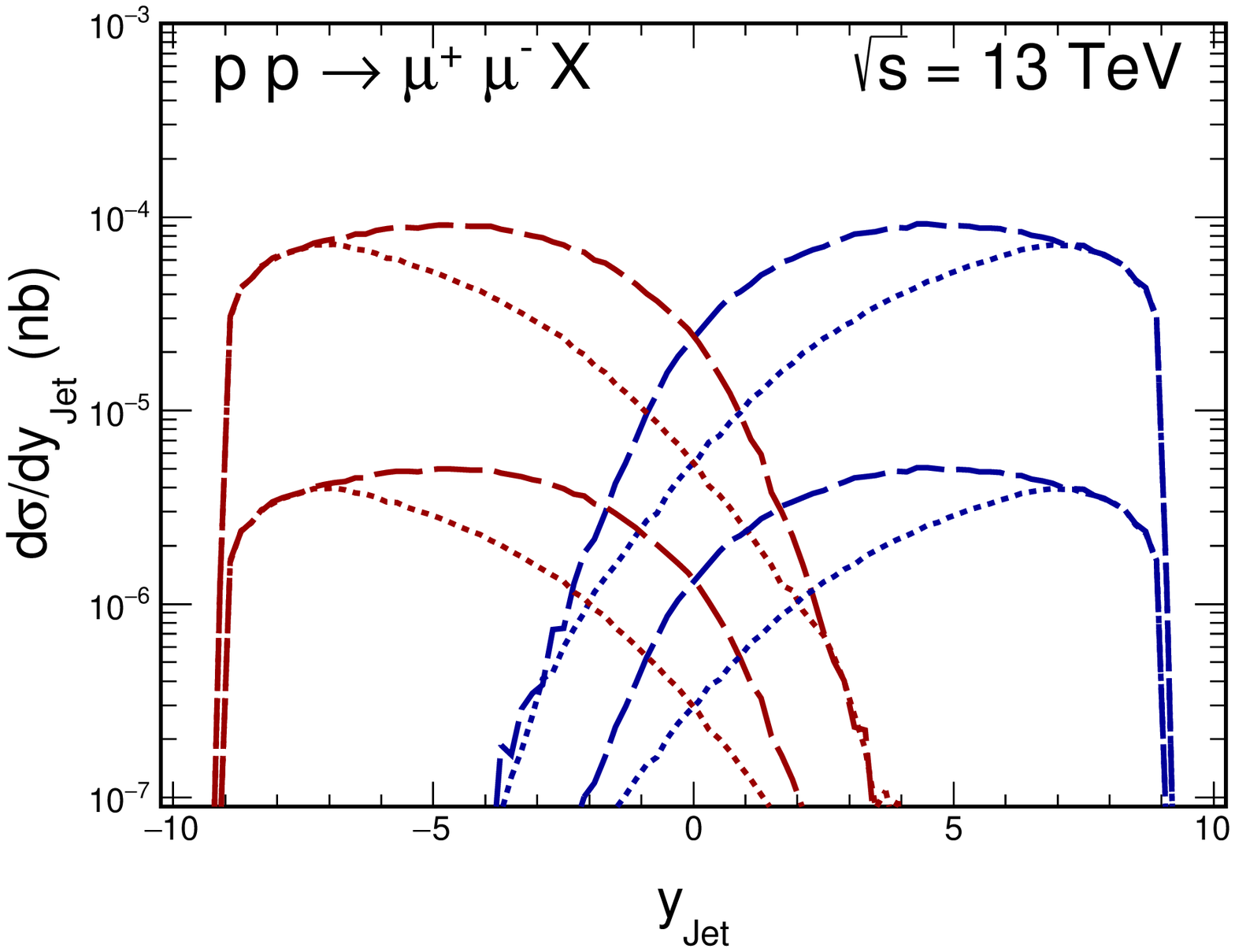}
\caption{Distribution in rapidity of (mini)jets for inclusive case
(upper curves) and for the case with cut on $\xi_{1/2}$ (lower curves).
The extra dotted lines represent results that include 
the cut $p_{t,pair} <$ 5 GeV as in the ATLAS analysis \cite{ATLAS}.}
\label{fig:dsig_dyjet}
\end{figure}
%------------------------------------------------------------------

\begin{figure}
\includegraphics[width=7cm]{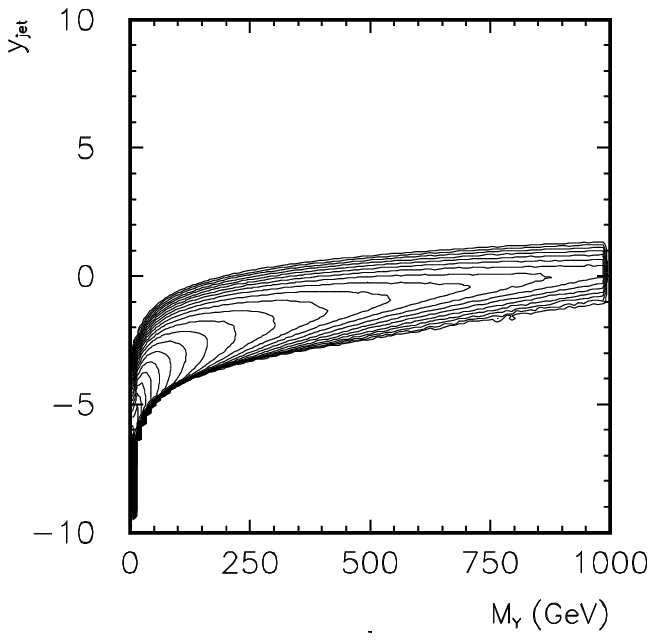}
\caption{Remnant mass - jet rapidity correlations.
The standard $\xi$-cuts were applied here.}
\label{fig:map_MRyjet}
\end{figure}

The emitted jet when entering the main detector (ATLAS or CMS) 
will destroy rapidity gap and such cases are usually vetoed in experiment.
Here we wish to show what are correlations of jet rapidity with
mass of the remnant. In Fig.\ref{fig:map_MRyjet} we
show such correlations for inelastic-elastic (left panel) and
elastic-inelastic (right panel) contributions.
If the mass of the remnant system is small the corresponding jet
is emitted outside of the main detector and is therefore not observed.
On the contrary the jets corresponding to large masses enter the main
detector and destroy the rapidity gap.

%------------------------------------------------------------------
%------------------------------------------------------------------

%Finally in Fig.\ref{fig:dsig_dptsum_withcuts} we show distribution in
%the lepton pair transverse momentum.
%The distributions for elastic-elastic components are very steep
%compared to those for the single-dissociative contributions.\\
%{\bf The two dissociative contributions give slightly different
%  distributions in $p_{t,sum}$ which must be understood.}

%--------------------------------------------------------------
%\begin{figure}
%\includegraphics[width=6cm]{dsig_dptsum_xillcuts.eps}
%\caption{Distribution in dilepton transverse momentum for
%the four different contributions considered.
%Here the cuts on $\xi_{ll}^{+}$ or $\xi_{ll}^-$ are imposed.}
%\label{fig:dsig_dptsum_withcuts}
%\end{figure}
%--------------------------------------------------------------

%{\bf So far we do not understand the asymmetries for
%the both single-dissociative contributions.}

%--------------------------------------------------------------------------
\subsection{SuperChic results and gap survival factor}
%--------------------------------------------------------------------------

For comparison we did also calculations using a popular SuperChic program
\cite{HTKR2020}. In this code the soft gap survival factor can be
included. We wish to study how the effective gap survival factor changes
with implementation of the cut on $\xi_1$ or $\xi_2$.
In the SuperChic generator the gap survival is calculated with opacity
obtained within two-channel Good-Walker formalism. The parameters of 
the model were obtained in \cite{KMR2013}.
The results presented here were obtained with the event sample 
$N_{event}$ = 50000, unless otherwise stated.

The integrated cross sections are collected in Table II.
The cross sections obtained in SuperChic are somewhat larger than those 
obtained using our code (compare the numbers in Table II with the
numbers in Table I). 
This is partially due to different structure functions used in both
codes. The cuts on $\xi$ in this table were calculated as:
\begin{equation}
\xi_{1/2} = \frac{\sqrt{s}/2 - E_{3/4}}{\sqrt{s}/2} \; ,
\end{equation}
where $E_{3}$ or $E_{4}$ are energies of outgoing protons.

%---------------------------------------------------------------------
\begin{table}
   \centering
   \caption{Integrated cross section for $\mu^+ \mu^-$ production
            in pb for $\sqrt{s}$ = 13 TeV using 
   SuperChic program \cite{HTKR2020}.
   We show results without any extra external cuts (upper part),
%   \footnote{There is internal cut -2.5 $<  Y_{ll} <$ 2.5 in the
%     SUPERCHIC generator.},
   with extra cut on individual rapidities -2.5 $< y_1, y_2 <$ 2.5
   (middle part) and with extra cut on $\xi$ (lower part).     
   The cut on $\xi_i$ is:  0.035 $ < \xi_{ll}^{\pm} < $ 0.08.
   In addition in this calculation $p_{1t}, p_{2t} >$ 15 GeV.
   We show result with (first column) and without (second column)
   soft gap survival. To calculate absorption effects we used 
   model no 4 as implemented in the SuperChic generator 
   (see also \cite{KMR2013}).
   In the last column we show average gap survival factor being the ratio
   of the cross sections in previous two columns.
   In all cases $p_{1t}, p_{2t} >$ 15 GeV.
   The numbers with (*) in the last block were obtained with 10 000
   events only. } 
\begin{tabular}{|c|c|c|c|}
\hline
reaction &  no soft $S_G$   &  with soft $S_G$ & $<S_G>$  \\
\hline
-2.5 $< Y_{ll} <$ 2.5  &  &  & \\
\hline
elastic-elastic                   &  0.54438  &  0.50402  &  0.926 \\   
inelastic-elastic                 &  0.89595  &  0.64283  &  0.717 \\
elastic-inelastic                 &  0.89587  &  0.64254  &  0.717 \\
inelastic-inelastic               &  1.62859  &  0.24172  &  0.148 \\
\hline
-2.5 $< y_1, y_2 <$ 2.5 in addition  &  &  & \\
\hline
elastic-elastic                   &  0.42268  &  0.39355 &  0.931 \\
inelastic-elastic                 &  0.69241  &  0.51092 &  0.738 \\
elastic-inelastic                 &  0.69246  &  0.51087 &  0.738 \\
\hline
$\xi$ cut in addition &  &  &  \\
\hline
elastic-elastic, cut on $\xi_1$   &  0.00762  &  0.00675  &  0.886 \\
elastic-elastic, cut on $\xi_2$   &  0.00762  &  0.00675  &  0.886 \\
%inelastic-elastic, cut on $\xi_2$ &  0.02718  &  0.01416  &  0.521 \\
%elastic-inelastic, cut on $\xi_1$ &  0.02717  &  0.01416  &  0.521 \\
inelastic-elastic, cut on $\xi_2$ & 0.02496 & 0.01324 &  0.530 \\
elastic-inelastic, cut on $\xi_1$ & 0.02393 & 0.01238 &  0.517 \\
\hline
$p_{t,pair} <$ 5 GeV in addition  &  &  &  \\
\hline
elastic-elastic                   &  .......  &  .......  &  ..... \\
%inelastic-elastic, cut on $\xi_2$ &  0.008056 (2000) &  0.00435 &  0.541 \\
%elastic-inelastic, cut on $\xi_1$ &  0.008035 (2000) &  0.00436 &  0.541 \\
inelastic-elastic, cut on $\xi_2$ &  0.00807 &  0.00437 (*) &  0.541 \\
elastic-inelastic, cut on $\xi_1$ &  0.00807 &  0.00437 (*) &  0.542 \\
\hline
\end{tabular}
\label{table:cs_with_xicuts_SUPERCHIC}
\end{table}
%------------------------------------------------------------------------

The typical cuts on $\xi_1$ or $\xi_2$ lower the cross section by almost 
two orders of magnitude. The numbers obtained here (SuperChic) after
the $\xi$ cuts included are, however, significantly larger than 
their conterparts in Table I.
%{\bf We do not understand this difference.}
The soft gap survival factor strongly depends
on whether we have fully elastic or single dissociation process.
This is related to typical transverse momenta of outgoing protons
which are bigger for processes with proton dissociation.
Similar tendencies can be observed when cuts on fractional 
longitudinal momentum loss of protons is imposed. Then, however, 
the soft gap survival factors are significantly reduced 
(0.931 $\to$ 0.886 for double elastic contribution and 
0.738 $\to$ 0.521 for processes with single dissociation).
The effect of explicit cuts on $y_1$ and $y_2$ is small and leads
to slight increase compared to the case of an internal cut on $Y_{ll}$
imposed by deafult in SuperChic.

Now we shall discuss the effect of gap survival factor on differential
distributions. In this case we shall use event sample generated by
the SuperChic 4 generator. For this purpose we generated 
5 10$^{+4}$ events and written a simple code which prepares
distributions of interest.
The statistics is rather small so the differential distributions will
fluctuate much more in comparison to distributions obtained from our codes
based on the VEGAS algorithm.

\begin{figure}
\includegraphics[width=7cm]{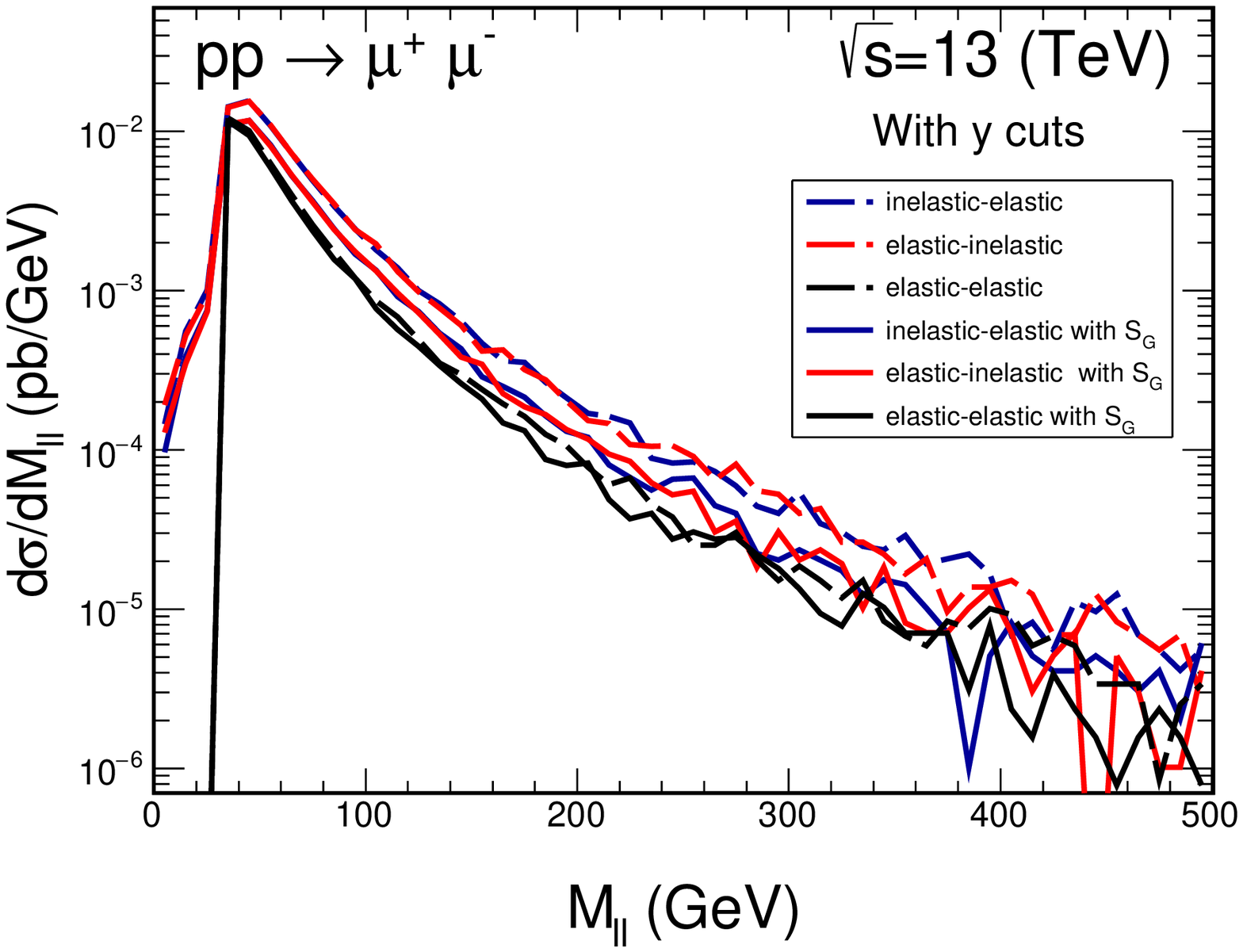}
\includegraphics[width=7cm]{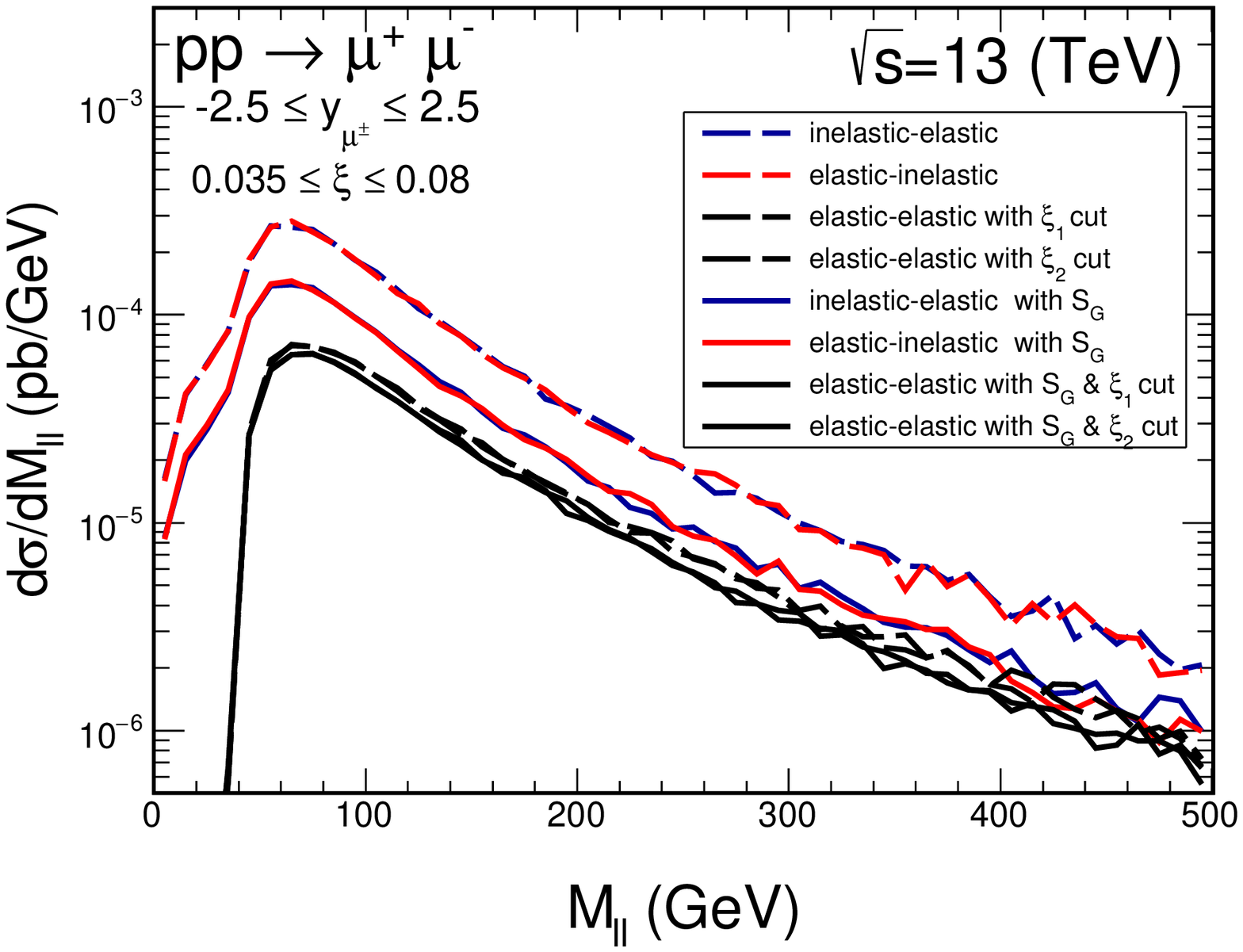}
\caption{Distribution in dimuon invariant mass for
the different contributions considered.
We consider the case without $\xi$ cuts (left panel) and
with $\xi$ cuts (right panel).
}
\label{fig:dsig_dMll_SUPERCHIC}
\end{figure}

In Fig.\ref{fig:dsig_dMll_SUPERCHIC} we present distribution 
in dimuon invariant mass for the case without $\xi$ cuts (left panel)
and with $\xi$ cuts (right panel). The elastic-elastic (dashed line)
and elastic-inelatic + inelastic-elastic (solid line) are shown
separately.
On average we observe larger invariant masses in the case with $\xi$ cuts.
%..................................................................\\
%..................................................................\\
%..................................................................\\

%--------------------------------------------------------------
%--------------------------------------------------------------

\begin{figure}
\includegraphics[width=7cm]{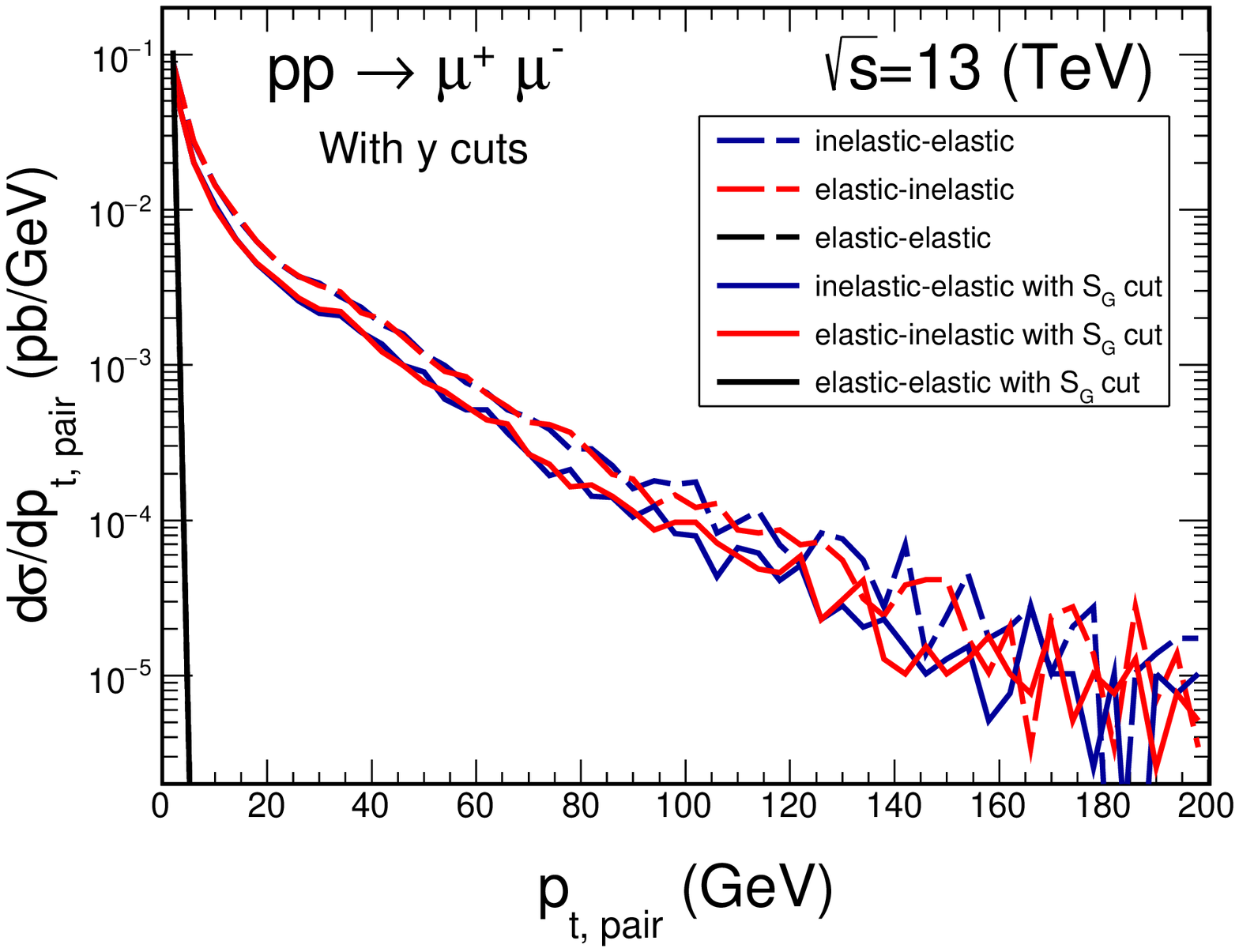}
\includegraphics[width=7cm]{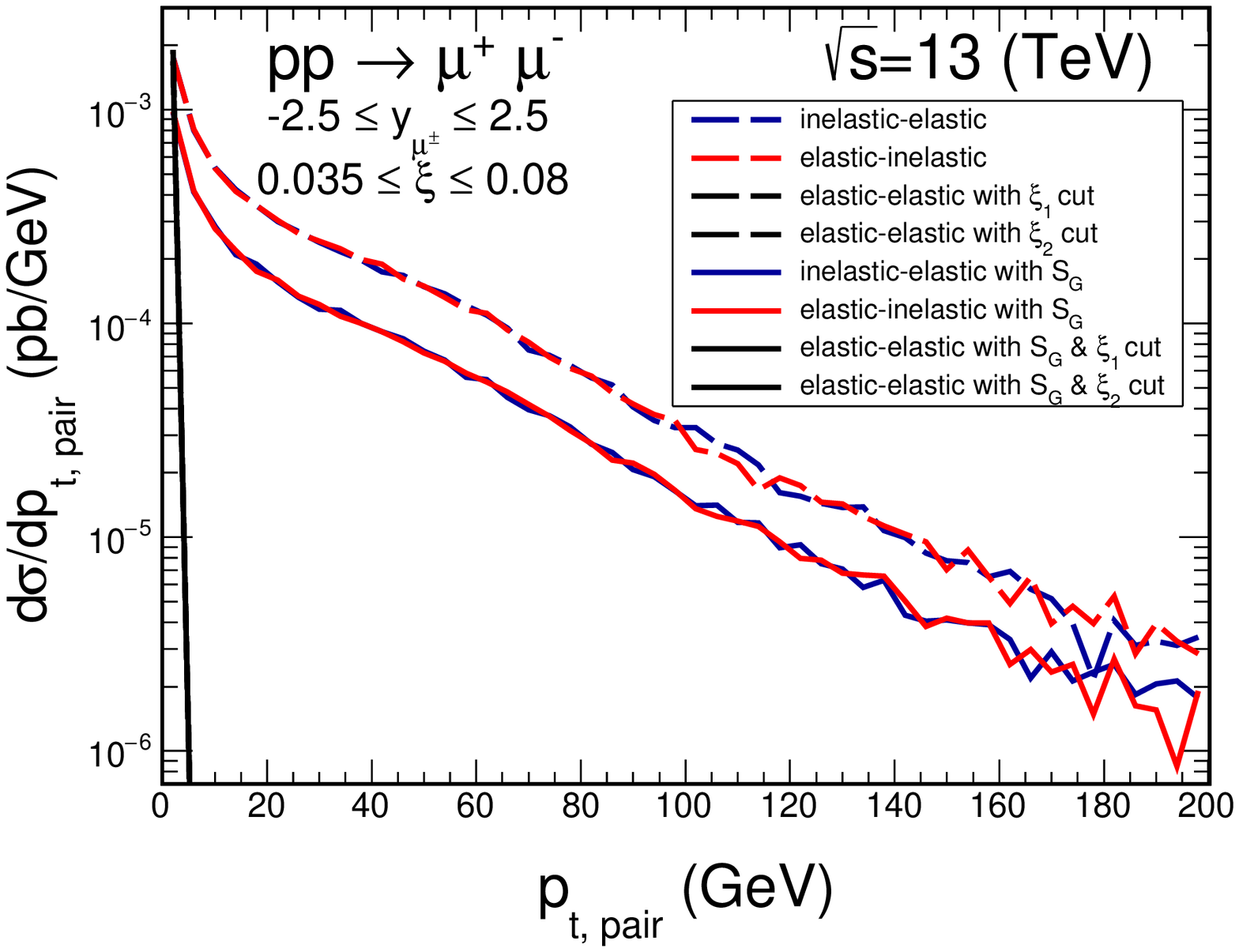}
\caption{Distribution in dimuon transverse momentum for
the different contributions considered.
We consider the case without $\xi$ cuts (left panel) and
with $\xi$ cuts (right panel).
}
\label{fig:dsig_dptsum_SUPERCHIC}
\end{figure}

In Fig.\ref{fig:dsig_dptsum_SUPERCHIC} we present similar 
distributions but in $p_{t,pair}$.
The shapes of distributions obtained with and without soft gap survival
effects seems rather similar.
%..................................................................\\
%..................................................................\\
%..................................................................\\

%--------------------------------------------------------------
%--------------------------------------------------------------

\begin{figure}
\includegraphics[width=7cm]{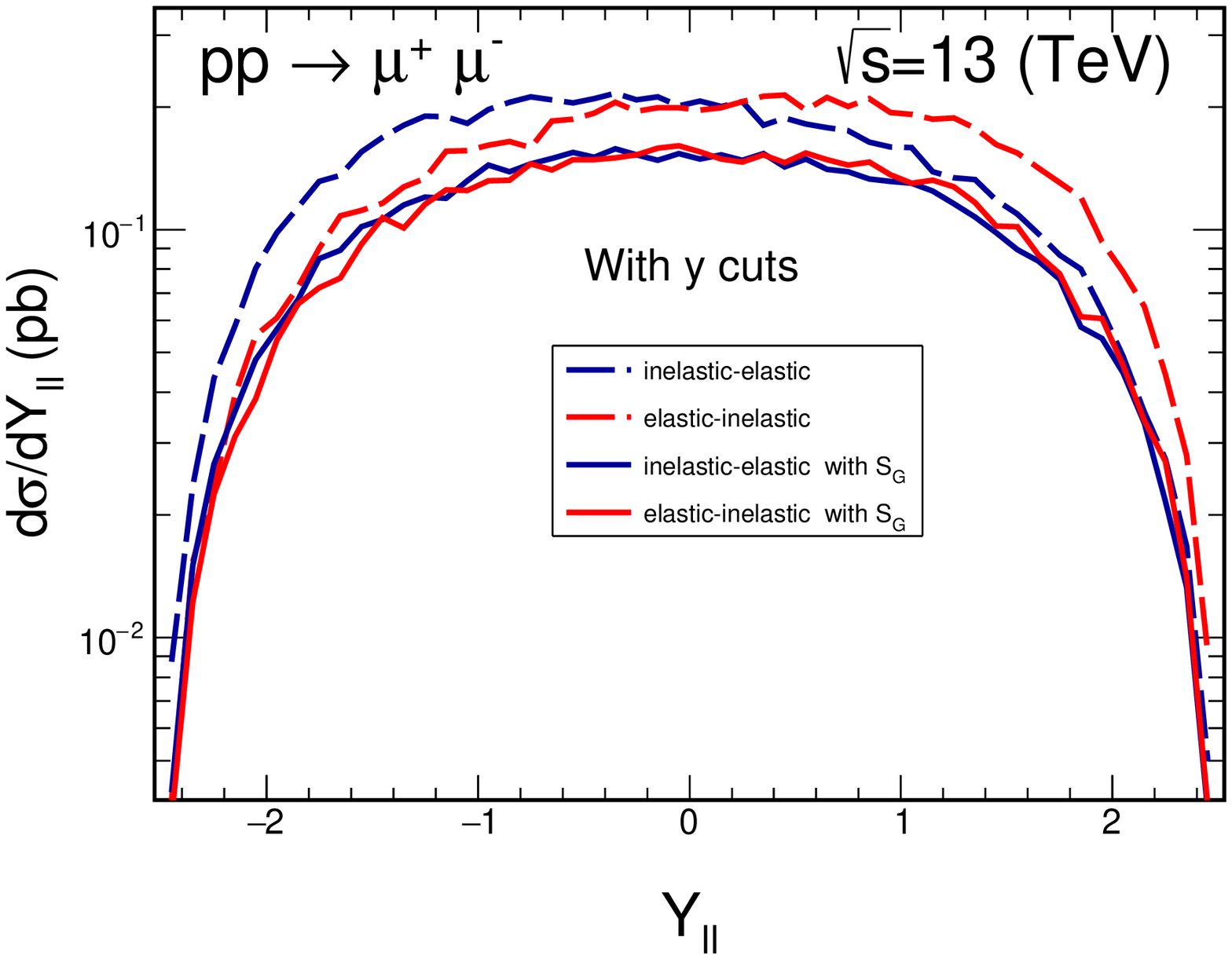}
\includegraphics[width=7cm]{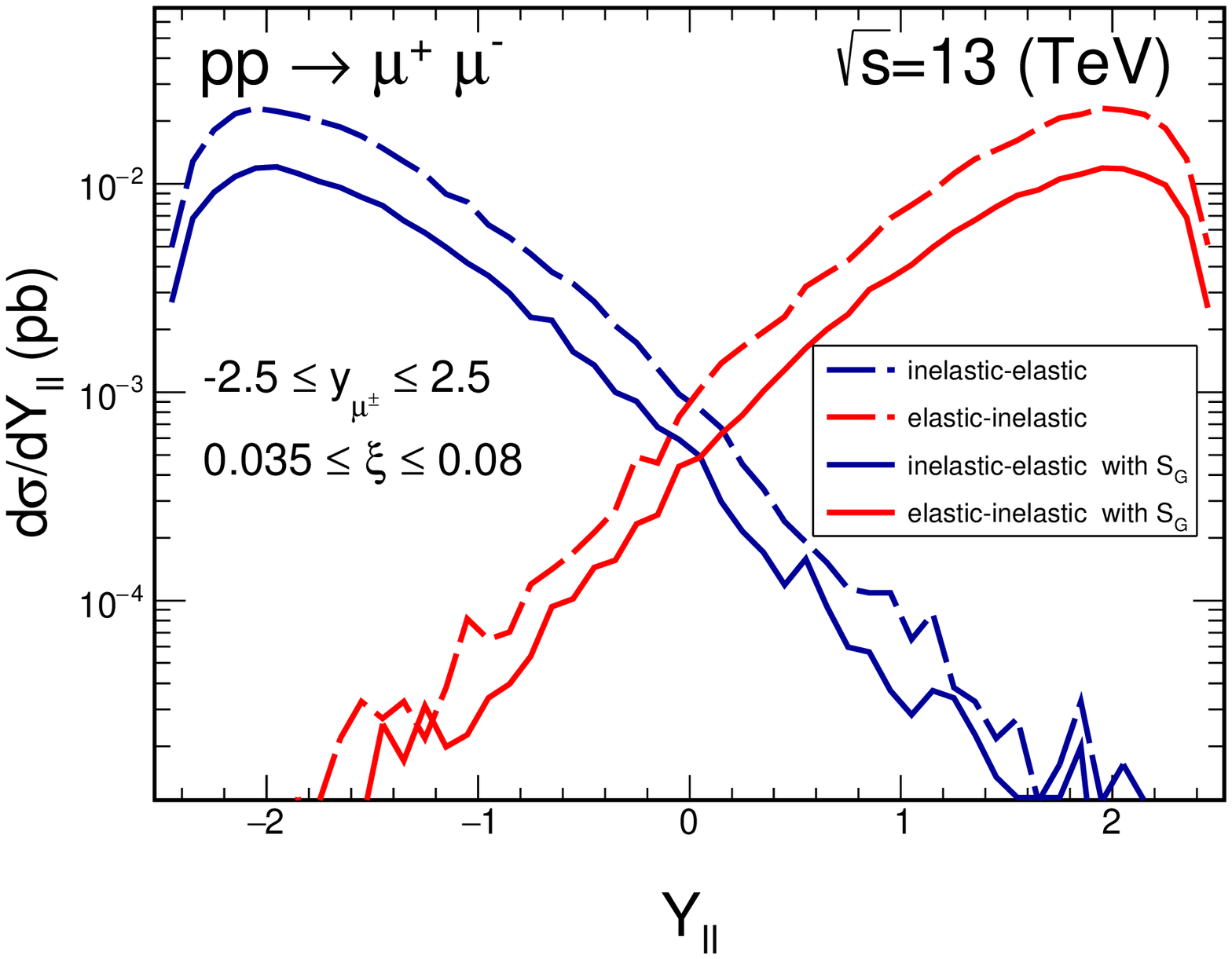}
\caption{Distribution in rapidity of the dimuon pair.
We show the case without $\xi$ cuts (left panel) and
with $\xi$ cuts (right panel).
for the different contributions considered.
}
\label{fig:dsig_dYll_SUPERCHIC}
\end{figure}

In Fig.\ref{fig:dsig_dYll_SUPERCHIC} we present similar
distributions but in $Y_{ll}$.
Without the $\xi$ cut we observe quite different shapes of distributions
in $Y_{ll}$ without and with soft rapidity gap survival factor 
(see the left panel).
%{\bf We do not understand this result.}
When the $\xi$-cut is imposed the distributions with and without
soft rapidity gap survival factor have very similar shapes.
Then, however, the elastic-inelastic and inelastic-elastic
contributions are well separated in $Y_{ll}$. The sum of both
contributions has a characteristic dip at $Y_{ll}$ = 0. 
This is the same as was discussed in the previous section.

%---------------------------------------------------------------
%---------------------------------------------------------------

\begin{figure}
\includegraphics[width=7cm]{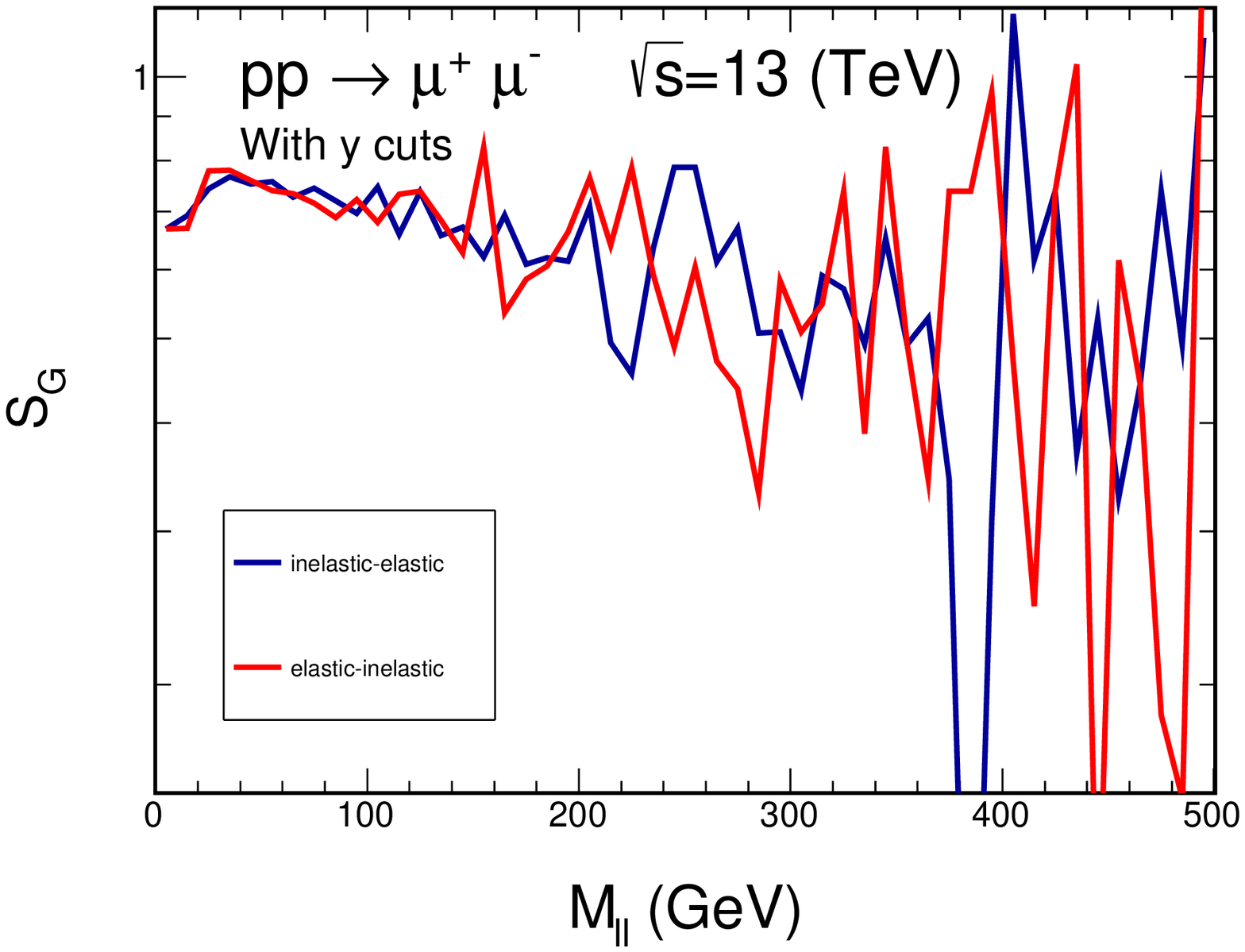}
\includegraphics[width=7cm]{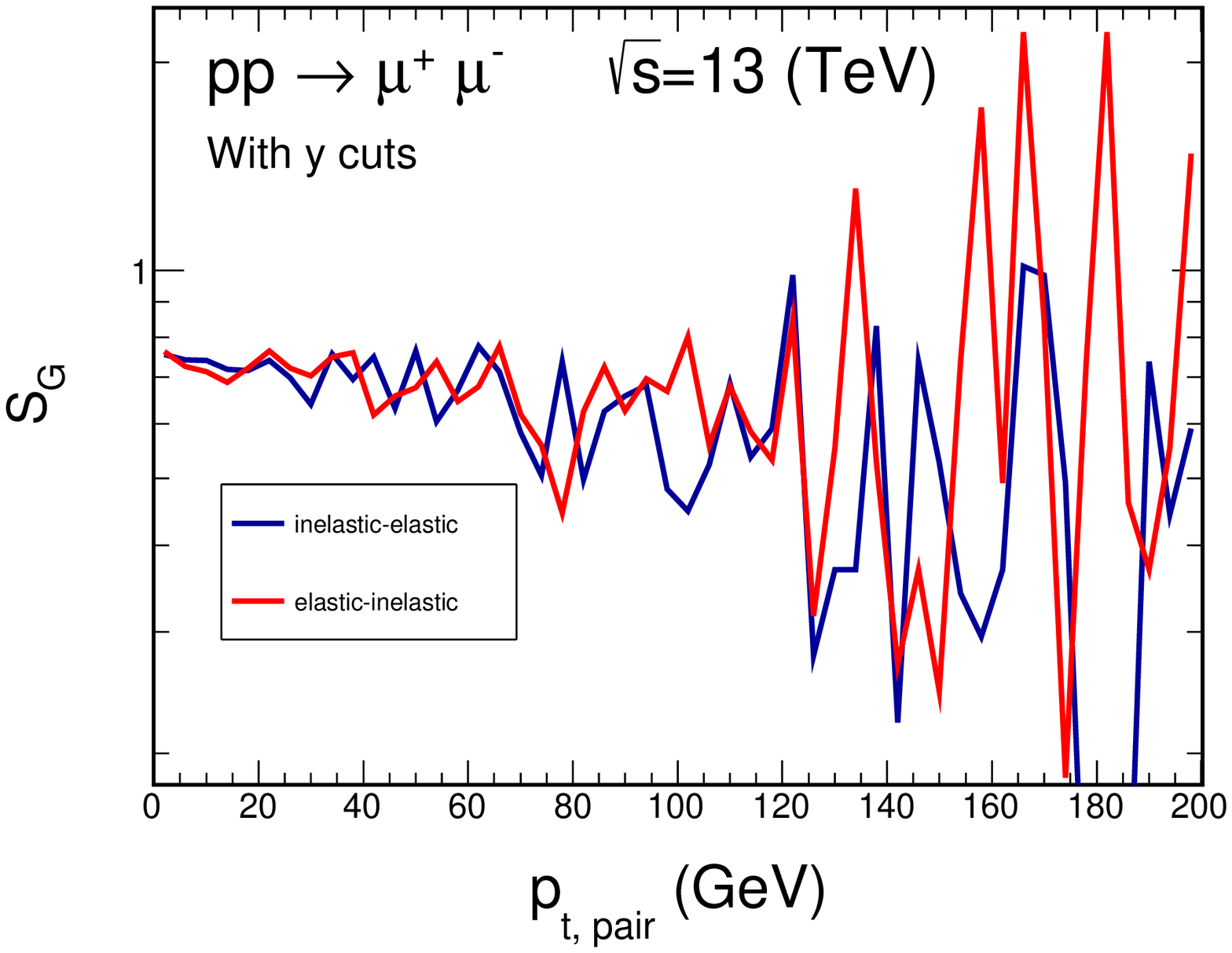}\\
\includegraphics[width=7cm]{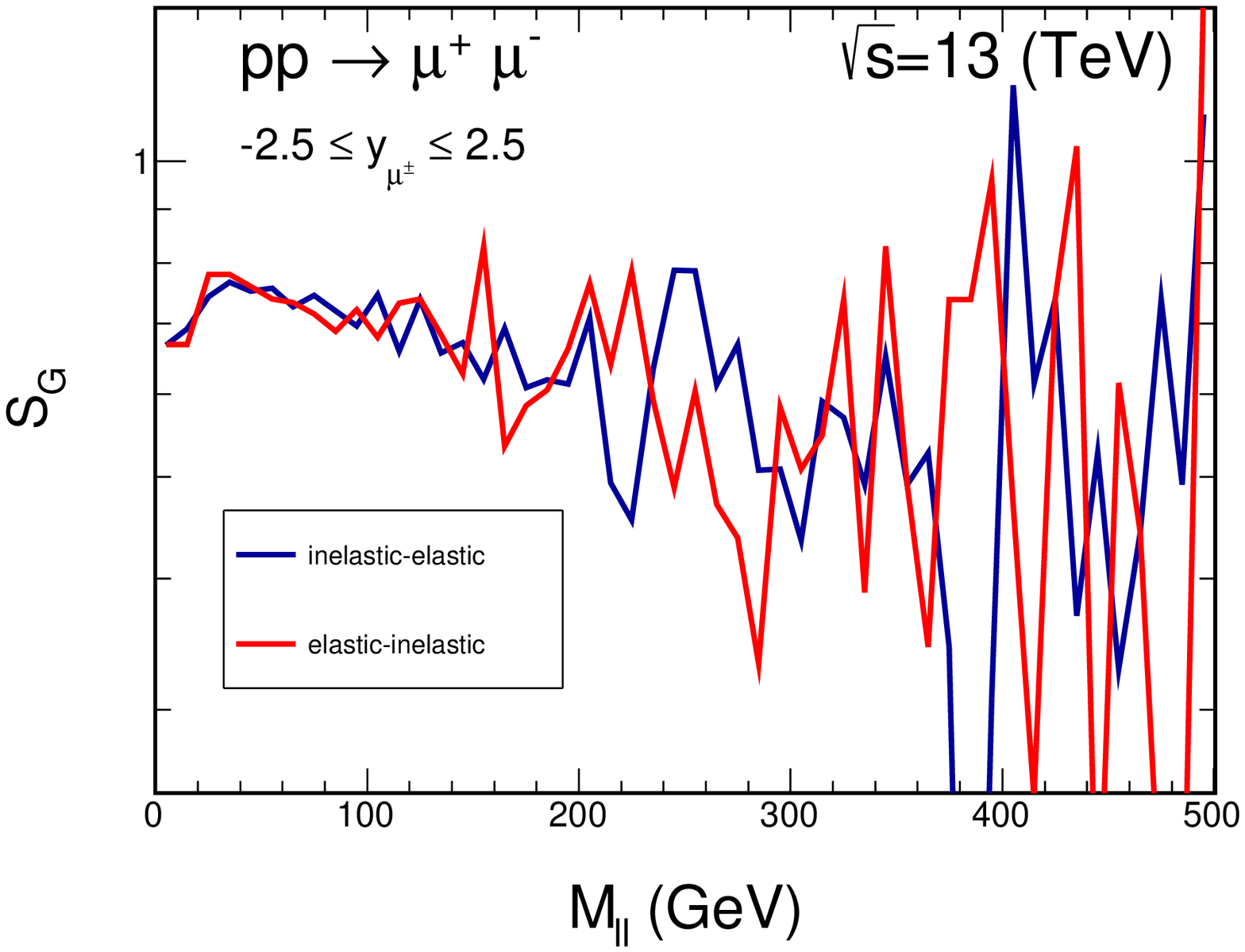}
\includegraphics[width=7cm]{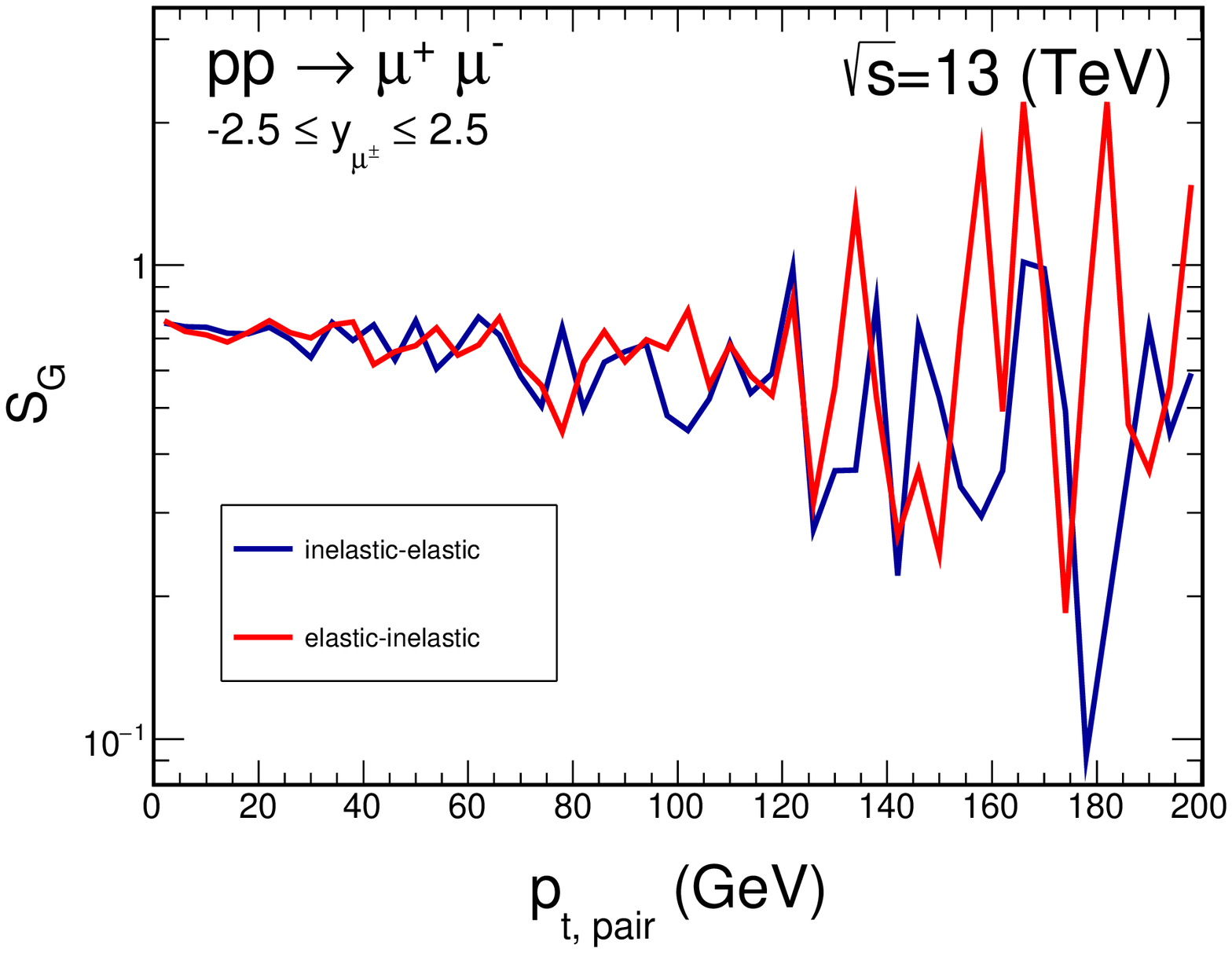}\\
\caption{The soft gap survival factor as a function of
dilepton invariant mass (left panels) and as a function of transverse
momentum of the pair (right panels) for 
%double elastic (dashed line) and 
single dissociation (solid line) mechanisms.
We show the result without $\xi$ cuts (upper panels) and
with $\xi$ cuts (lower panels).
}
\label{fig:soft_gap_survival_factor}
\end{figure}

In Fig.\ref{fig:soft_gap_survival_factor} we show corresponding
gap survival factor calculated as:
\begin{eqnarray}
S_G(M_{ll}) &=& \frac{d \sigma / d M_{ll}|_{with SR}}
                     {d \sigma / d M_{ll}|_{without SR}}
\; , \\
S_G(p_{t,pair}) &=& \frac{d \sigma / d p_{t,pair}|_{with SR}}
                         {d \sigma / d p_{t,pair}|_{without SR}}
\; ,
\label{differential_gap}
\end{eqnarray}
the ratio of the cross section with the soft rapidity gap survival
factor to its counterpart without including the effect,
differential in $M_{ll}$ (left panel)
or in $p_{t,pair}$ (right panel) for double elastic (dashed line)
and single dissociation (solid line).
We observe a small dependence on both $M_{ll}$ and on $p_{t,pair}$. 
The gap survival factor for double elastic component
is larger than that for single dissociation.
The gap survival factor corresponding to the measurement of one proton 
is significantly smaller than that for the inclusive case.
The rather large fluctuations are due to limited statistics (50 000 events).

%-------------------------------------------------------------------
%--------------------------------------------------------------------

\begin{figure}
\includegraphics[width=7cm]{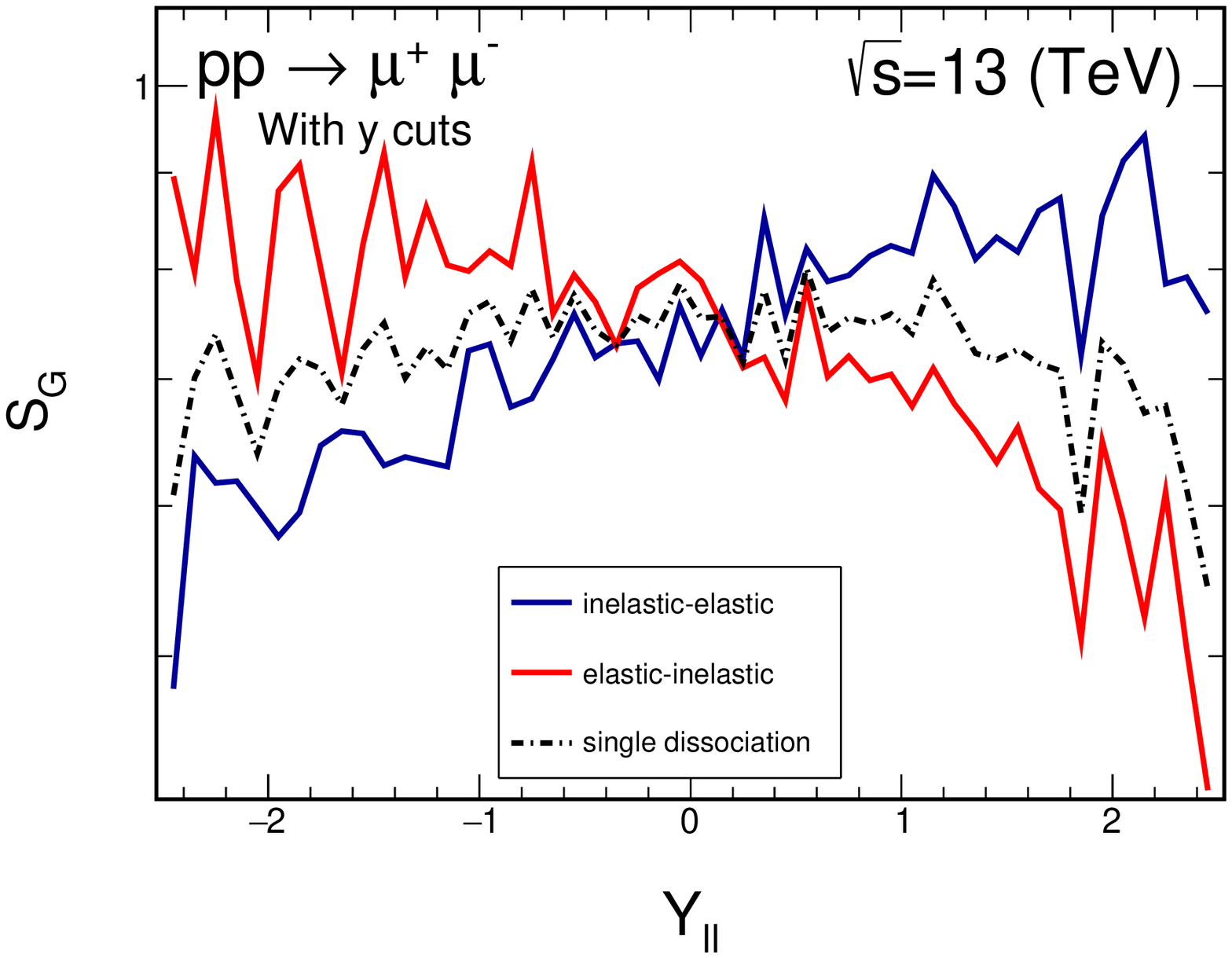}
\includegraphics[width=7cm]{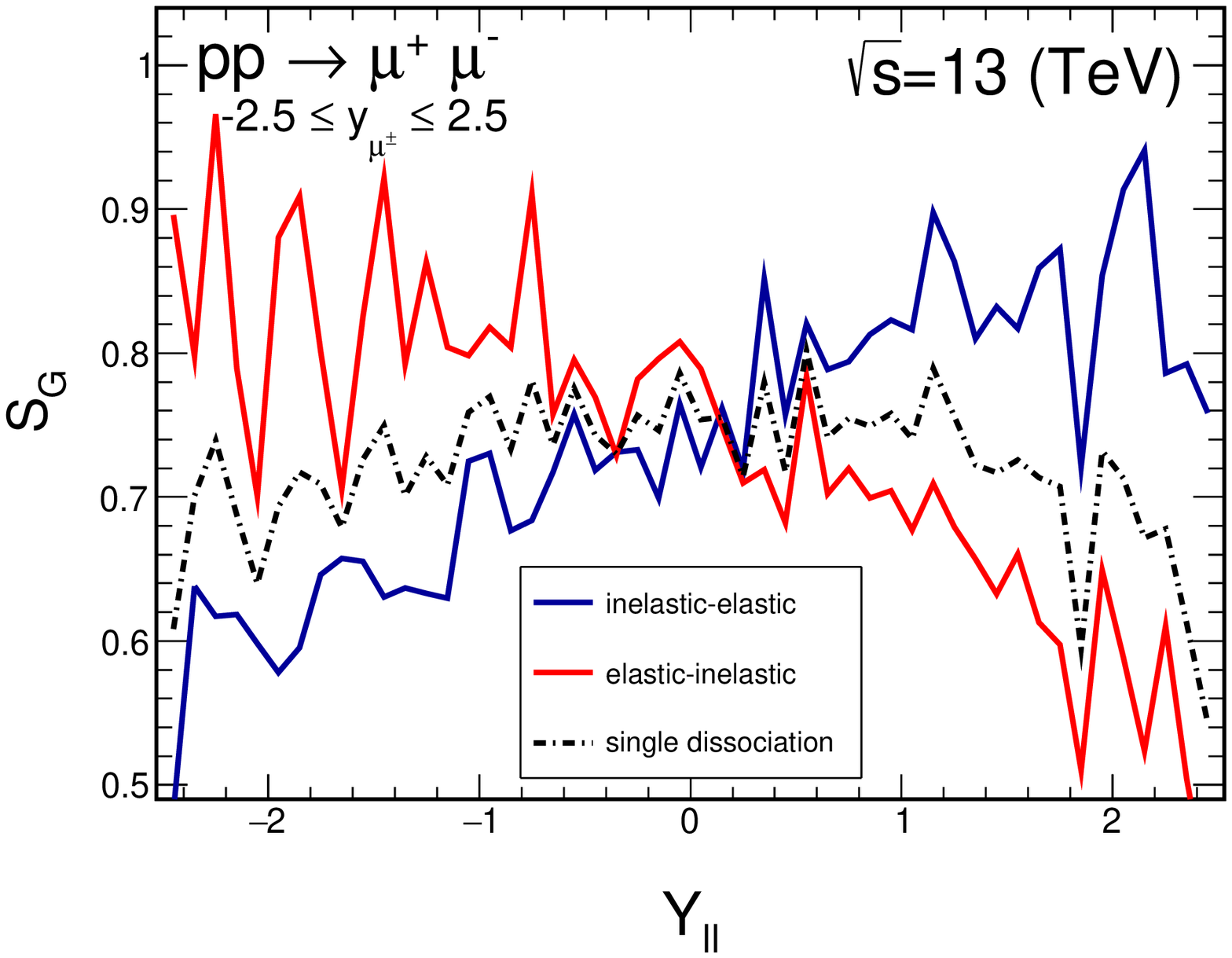}
\caption{The soft gap survival factor as a function of
rapidity of the $\mu^+ \mu^-$ pair for single proton dissociation.
We show the result without $\xi$ cuts (left panel) and
with $\xi$ cuts (right panel). The dash-dotted black line represents
effective gap survival factor for both single-dissociation components
added together.
}
\label{fig:soft_gap_survival_factor_2}
\end{figure}

In Fig.\ref{fig:soft_gap_survival_factor_2} we show in addition 
soft gap survival factor as a function of the rapidity of the dimuon pair.
We observe a strong dependence of the gap survival factor on $Y_{ll}$
separately for elastic-inelastic and inelastic-elastic components
but only in the case when proton is not measured. This effect may
be very difficult to address experimentally as in this 
(no proton measurement) case one measures the sum of the both (all)
components, where the effect averages and becomes more or less
independent of $Y_{ll}$ (see black dash-dotted curve).
However, it seems interesting to understand the dependence on $Y_{ll}$ 
for individual component from theoretical point of view.\\
%{\bf Disscusion is welcome.}

%---------------------------------------------------------------------
%---------------------------------------------------------------------

%Rysunki do zrobienia:\\

%1) dsigma/dMll      (a) gap vs no gap \\
%                    (b) no xi cut and xi cut \\

%2) dsigma/dptll     (a) gap vs no gap \\ 
%                    (b) no xi cut and xi cut \\

%3) dsigma/dyl+   , dsigma/dyl- \\

%.................................................................\\
%.................................................................\\
%.................................................................\\

\begin{figure}
\includegraphics[width=7cm]{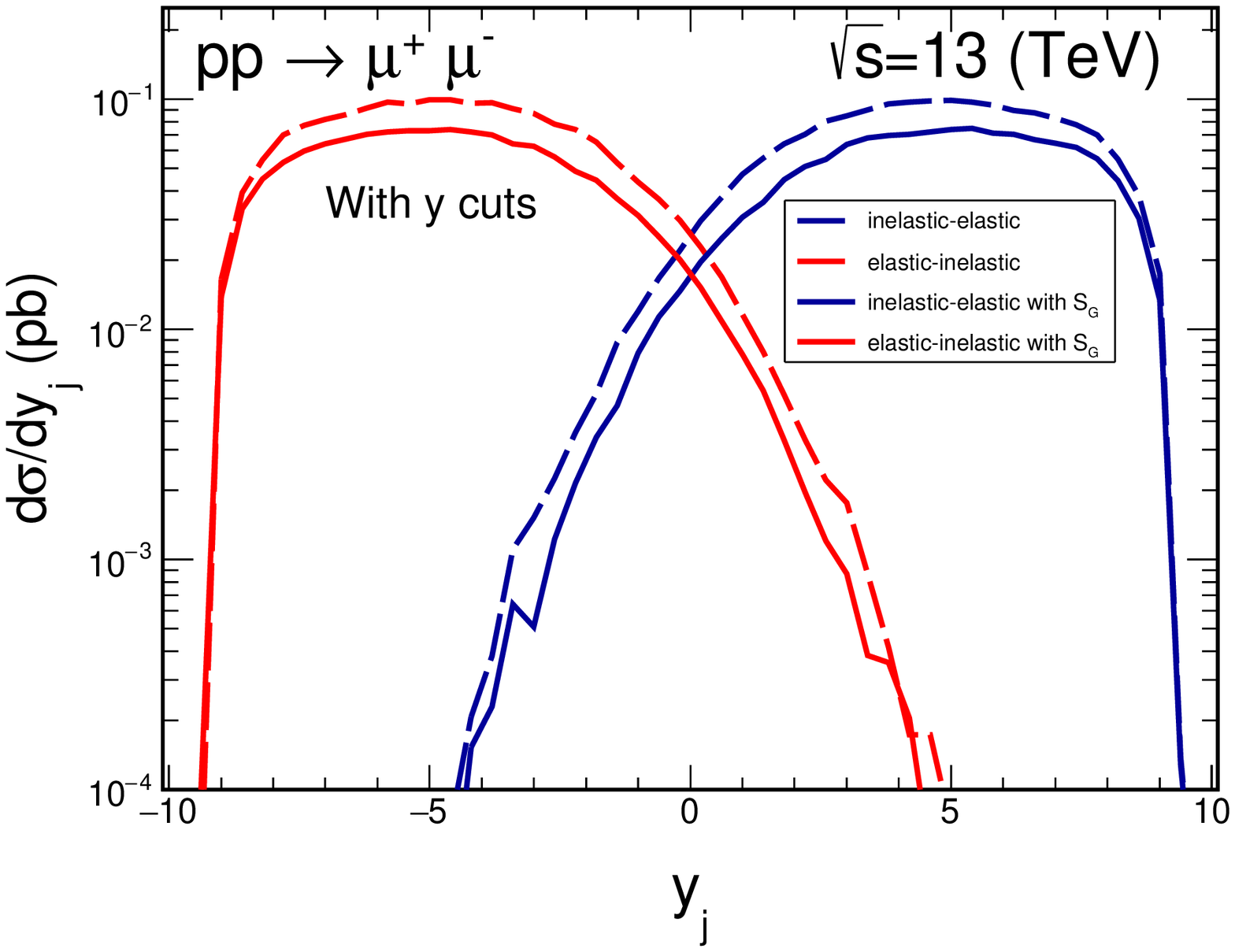}
\includegraphics[width=7cm]{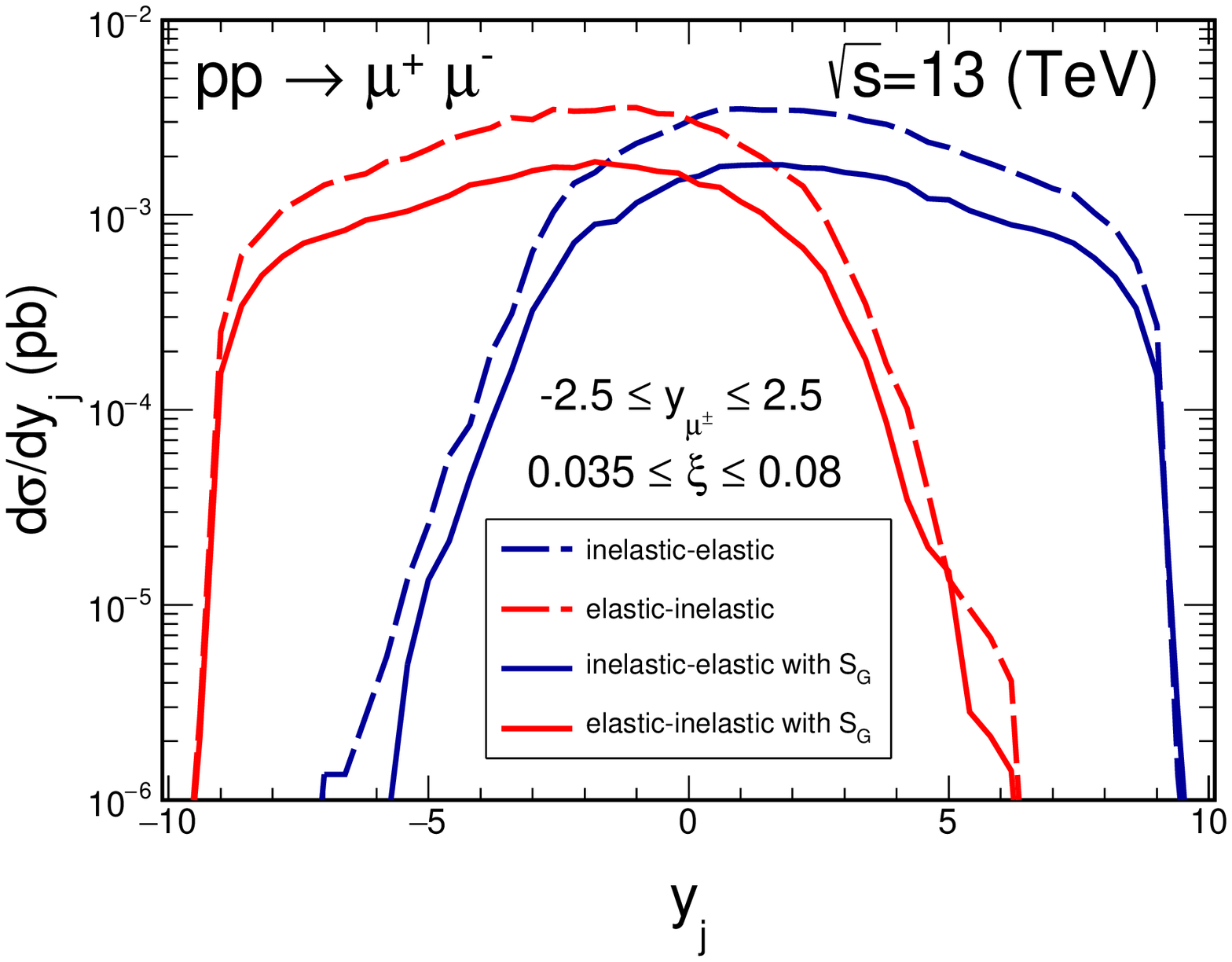}
\caption{Distribution in the (mini)jet rapidity for the inclusive case
with no $\xi$ cut (left panel) and when the cut on $\xi$ is imposed
(right panel) for elastic-inelastic and inelastic-elastic contributions
as obtained from the SuperChic generator.
We show result without (dashed line) and with (solid line) soft
rescattering correction.
}
\label{fig:dsig_dyjet_SUPERCHIC}
\end{figure}

Finally we wish to discuss how the proton dissociation further reduces
the gap survival factor due to emission of a (mini)jet that can enter
into the main detector and destroy the rapidity gap.
\footnote{However, the rapidity gap condition was not explicitly imposed
in recent analyses with forward proton measurements  
\cite{CMS} or \cite{ATLAS}.}
This was discussed e.g. in \cite{HKR2016,LSS2018,LFSS2019}.
In Fig.\ref{fig:dsig_dyjet_SUPERCHIC} we show the (mini)jet distribution
in rapidity for elastic-inelastic and inelastic-elastic components.
We show the distribution without imposing the $\xi$ cut (left panel)
and when imposing the $\xi$ cut (right panel).
One can observe slightly different shape for both cases.
The corresponding gap survival factor (probability of no jet in the main
detector) is 0.8 and 0.5, respectively.
The probability of no emission around the 
$\gamma \gamma \to \mu^+ \mu^-$ vertex is, however, much more difficult
to calculate and requires inclusion of remnant hadronization
which is model dependent.

%-----------------------------------------------------------------------
%-----------------------------------------------------------------------

In Table III we show probability that the (mini)jet is
outside the main detector, i.e.: $y_{jet} <$ -2.5 or $y_{jet} >$ 2.5.
The numbers below are similar in size to the soft gap survival factor
collected in Table II.
Imposing cuts on $\xi$ lowers the corresponding (minijet) rapidity 
gap survival factor while imposing extra cut $p_{t,pair} <$ 5 GeV, 
as in the ATLAS experiment, increases it back.

The factor below must be evidently included in the case when rapidity
gap condition is imposed experimentally. It is less clear what to do
when the condition of separated $\mu^+$ and $\mu^-$ are imposed
as in the ATLAS experiment \cite{ATLAS}.
In the following we assume that the particles from (mini)jet, emitted 
from the same vertex as leptons, will always break the conditions, 
provided they are emittted in the same range of rapidities as 
the measured leptons. 
This range is defined by the geometry of the main ATLAS (CMS) detector.
In real experiment \cite{ATLAS} one imposes rather condition
on $R(track,l^+) > R_0$ and $R(track,l^-) > R_0$ (no emission in the
cones around both leptons). In the ATLAS experiment $R_0$ = 0.01 was used.
It would be interesting to study experimentally the gap survival
factor as a function of $R_0$.\\
%{\bf More discussion is welcome.}

%---------------------------------------------------------------------
\begin{table}
   \centering
   \caption{Gap survival factor due to minijet emission.
The first block is with only internal SuperChic cut:
-2.5 $< Y_{ll} <$ 2.5, the second block is when the condition
on individual rapidities is imposed extra,
the third block includes in addition the cut on $\xi_1$ or $\xi_2$,
and the final block includes also the condition $p_{t,pair} <$ 5 GeV.
In all cases $p_{1t}, p_{2t} >$ 15 GeV.
In the last panel (*) means 10 000 events only.} 

\begin{tabular}{|c|c|c|}
\hline
contribution & without $S_G$ & with $S_G$ \\
\hline
cut on $Y_{ll}$ only &  & \\
\hline
elastic-inelastic & 0.76304 & 0.78756 \\
inelastic-elastic & 0.76278 & 0.78898 \\ 
\hline
cut on $y_1$ and $y_2$ in addition &  & \\
\hline
elastic-inelastic & 0.77366 & 0.79250 \\  
inelastic-elastic & 0.76926 & 0.78744 \\
\hline
cut on $\xi_1$ or $\xi_2$ in addition &   & \\
\hline
%elastic-inelastic & 0.48954 & 0.49986 \\
%inelastic-elastic & 0.48374 & 0.49508 \\
elastic-inelastic & 0.52430 & 0.53976 \\
inelastic-elastic & 0.53118 & 0.53614 \\
\hline
cut on $p_{t,pair}$ in addition &   &   \\
\hline
%elastic-inelastic & 0.8395 & 0.8600 \\
%inelastic-elastic & 0.8505 & 0.8560 \\ 
elastic-inelastic & 0.83144 & 0.84350(*) \\
inelastic-elastic & 0.83462 & 0.84960(*) \\ 
\hline
\end{tabular}
\label{table:jet-survival-factor}
\end{table}
%----------------------------------------------------------------------

%----------------------------------------------------------------------------
%\subsection{Gap survival factor for single-dissociative
%processes}
%----------------------------------------------------------------------------

%Particularly interesting is the gap survival factor for the case
%of at least one inelastic excitation.
%In this case we should study dependence of the gap survival factor
%on $M_X$ (elastic-inelastic) or $M_Y$ (inelastic-elastic) as well as 
%on $M_{ll}$.

%{\bf Marta should calculate the gap survival factor
%in the same way it as was done for the $W^+ W^-$ production.} \\

%-------------------------------------------
%\subsection{$\Delta^+$ contribution}
%-------------------------------------------

%{\bf Piotr could calculate it.}\\
%Alternatively we can include only Delta resonance in the Fiore et
%al. fit \cite{Fiore}.

%--------------------------
\section{Conclusions}
%--------------------------

In the present paper we have discussed dilepton production 
via photon-photon fusion with one forward proton which can 
be measured in forward detectors such as AFP for the ATLAS experiment.
We have considered both double-elastic and single-dissociative contributions
(it was argued that the contribution of double dissociation is 
negligible when forward proton is measured).
In the latter case we have considered both continuum production 
as well as $\Delta^+$ isobar production or production of other 
nucleon resonances. 
The continuum contribution is calculated for different parametrizations
of the deep-inelastic structure functions from the literature.
The differences of the cross section are of the order of 10-20 \% and
can be regarded as uncertainties of the present modelling.

We have imposed conditions on $\xi_1$ or $\xi_2$ for the forward 
emitted protons.
Several distributions have been shown and discussed in this case.
Particularly interesting is the distribution 
%in $M_{ll}$ and the distribution 
in $Y_{ll}$ which has a minimum at $Y_{ll} \sim$ 0.
The minimum at $Y_{ll}$ = 0 is caused by the experimental condition on 
$\xi_{ll}^{\pm}$ imposed on the leading proton.

We have also quantified the region of $x_{Bj}$ (the argument of the structure
functions) relevant for the $p p \to l^+ l^-$ processes. We found
that the typical values of $x_{Bj}$ are rather larger than 10$^{-3}$ both
for inclusive case and in the case with proton measurement.

We have also made calculations with the popular SuperChic generator 
and compared corresponding results to the results of our code(s).
In general, the results are very similar.

We have also calculated soft rapidity gap survival factor (probability
of no hadron emission in the range of the main (ATLAS, CMS) detector)
as a function of $M_{ll}$, transverse momentum of the dilepton pair,
mass of the proton remnant and $Y_{ll}$.
No evident {\bf dependencies} on the variables have been found
for the single dissociation, except of distribution in $Y_{ll}$.
We have found different (much larger) gap survival factor for fully elastic
contribution than for single proton dissociation.
The soft gap survival factor for single dissociative contribution 
strongly depends on whether proton is measured or not. 
It is significantly smaller when the forward proton is measured.

We have performed analysis of the range of arguments of the structure
functions relevant for photon-photon processes.
The photon virtualities are both in perturbative and nonperturbative
regions. The nonperturbative region becomes relatively larger when
the cut $p_{t,pair} <$ 5 GeV, as imposed in the recent ATLAS analysis
\cite{ATLAS}.

We have also calculated gap survival factor due to mini(jet) emission
by checking whether the minijet enters or not the main detector.
The second type of the gap survival also strongly depends on
whether the outgoing proton is measured or not. It is about 0.8 for 
inclusive case (no proton measurement) and about 0.5 for the case with 
proton measurement in the forward proton detector (with typical limited 
$\xi$ values).
The second type of gap survival factor is to large extent independent
of the soft gap survival factor, so in general, the two factors can be 
included multiplicatively. In our opinion it is not clear, however, what
to do in the case when lepton isolation cuts with specific parameters 
are included. Then inclusion of both effects multiplicatively
may lead to underestimation of the measured cross section.

In the present paper we have intentionally concentrated on discussing 
effects related to a measurement of one forward proton i.e. on imposing
$\xi$ cuts and absorption effect and not on direct comparison
to the new ATLAS data.

\vskip+5mm
{\bf Acknowledgments}\\

This study was partially supported by the Polish National Science Center
grant UMO-2018/31/B/ST2/03537 and by the Center for Innovation and
Transfer of Natural Sciences and Engineering Knowledge in Rzesz{\'o}w.
We are indebted to Jesse Liu, Rafa{\l} Staszewski and Marek Tasevsky
for a discussion and explanation of details of the recent ATLAS paper.

%----------------------------------------------------------------------------

%----------------------------------------------------------------------------

\end{document}